\documentclass{article}

\usepackage{arxiv}

\usepackage{graphicx}                % allow us to embed graphics files
\graphicspath{{figures/}{pictures/}{images/}{./}} % where to search for the images
\DeclareGraphicsExtensions{.eps}     % for pure latex we expect eps files

\usepackage{microtype}                 % use micro-typography (slightly more compact, better to read)
\PassOptionsToPackage{warn}{textcomp}  % to address font issues with \textrightarrow
\usepackage{textcomp}                  % use better special symbols
\usepackage{mathptmx}                  % use matching math font
\usepackage{times}                     % we use Times as the main font
\usepackage{cite}                      % needed to automatically sort the references
\usepackage{tabu}                      % only used for the table example
\usepackage{booktabs}                  % only used for the table example
\usepackage{acronym}
\usepackage{siunitx}
\usepackage{multirow}
\usepackage{multicol}
\usepackage{amsmath}
\usepackage{caption}
\usepackage{amssymb}
\usepackage{lineno}
\usepackage{lineno,microtype}
\usepackage[onehalfspacing]{setspace}
\usepackage[english]{babel}
\usepackage{graphicx}
\usepackage{stackengine}
\usepackage{comment}
\usepackage{enumitem}
\usepackage[compact, noindentafter]{titlesec}
\usepackage{float}
\usepackage{epstopdf}

\usepackage{subfig}

\def \figtwoscale {0.3}
\def \figthreescale {0.25}

\def\Effect{\mathit{Effect}}
\def\EffectorToTool{\mathit{EffectorToTool}}
\def\ToolToEffector{\mathit{ToolToEffector}}

\title{Docking Haptics: Extending the Reach of Haptics by Dynamic Combinations of Grounded and Worn Devices}

\author{Anthony Steed\thanks{ a.steed@ucl.ac.uk}, Sebastian Friston, Vijay Pawar, David Swapp\\
	\scriptsize Department of Computer Science, University College London, United Kingdom\\
}

\acrodef{ve}[VE]{Virtual Environment}
\acrodef{vr}[VR]{Virtual Reality}
\acrodef{ris}[RIS]{Real-time Interactive System}
\acrodef{dve}[DVE]{Distributed Virtual Environment}
\acrodef{cve}[CVE]{Collaborative Virtual Environment}
\acrodef{mvr}[MVR]{Mobile Virtual Reality}
\acrodef{rtt}[RTT]{Round-Trip Time}
\acrodef{qoe}[QoE]{Quality of Experience}
\acrodef{snr}[SNR]{Signal to Noise Ratio}
\acrodef{psd}[PSD]{Power Spectral Density}
\acrodef{rmse}[RMSE]{Root Mean Squared Error}
\acrodef{dof}[DOF]{Degree of Freedom}
\acrodefplural{dof}[DOFs]{Degrees of Freedom}
\acrodef{hci}[HCI]{Human Computer Interaction}
\acrodef{dip}[DIP]{Distal Interphalangeal}
\acrodef{pip}[PIP]{Proximal Interphalangeal}
\acrodef{mcp}[MCP]{Metacarpal}
\acrodef{cmc}[CMC]{Carpometacarpal}

\begin{document}

\maketitle

\begin{abstract}
Grounded haptic devices can provide a variety of forces but have limited working volumes. Wearable haptic devices operate over a large volume but are relatively restricted in the types of stimuli they can generate.
We propose the concept of docking haptics, in which different types of haptic devices are dynamically docked at run time. This creates a hybrid system, where the potential feedback depends on the user's location. We show a prototype docking haptic workspace, combining a grounded six degree-of-freedom force feedback arm with a hand exoskeleton. We are able to create the sensation of weight on the hand when it is within reach of the grounded device, but away from the grounded device, hand-referenced force feedback is still available. A user study demonstrates that users can successfully discriminate weight when using docking haptics, but not with the exoskeleton alone. Such hybrid systems would be able to change configuration further, for example docking two grounded devices to a hand in order to deliver twice the force, or extend the working volume. We suggest that the docking haptics concept can thus extend the practical utility of haptics in user interfaces.
\end{abstract}

\section{Introduction}
\label{sec:intro}

Kinaesthetic feedback devices are well studied in human computer interaction. The ability to exert forces against the user has important applications in areas such as tele-operation, rehabilitation and surgical simulation. With the advent of consumer \ac{vr}, there is a resurgence of interest in replacing the `missing' sensations of haptic feedback from virtual worlds. 
The generation of force feedback is challenging due to the conflicting requirements imposed on the devices. Devices must be large and stiff to render forces with high fidelity across large working volumes. However, large devices made of stiff materials are typically expensive, difficult to back-drive, and slower to respond. In \ac{vr} it is desirable to generate force feedback wherever the user is currently operating, but large and stiff devices typically have to be grounded due to their mass.
%This suggests only working with devices that the user can wear or hold, but these cannot generate net forces on the user to simulate, say, weight of an object. 

We propose the concept of {\em docking haptics} - a class of hybrid system that enables the dynamic re-configuration of different types of haptic devices to extend their workspace and capabilities. 
We suggest that by combining wearables or hand-helds with grounded devices we can extend the system capabilities beyond those of the individual devices and provide plausible feedback in a broader range of immersive simulations.

For example, a problem with worn devices such as hand exoskeletons is that while they can provide force feedback, it is only referenced to a nearby point on the body. Users can feel forces between their fingers \& wrist or palm, but not the strain in their muscles or tendons in the arm, torso or legs, leaving them unable to get a sensation of weight.

\begin{figure*}[h]
  \includegraphics[width=0.45\textwidth]{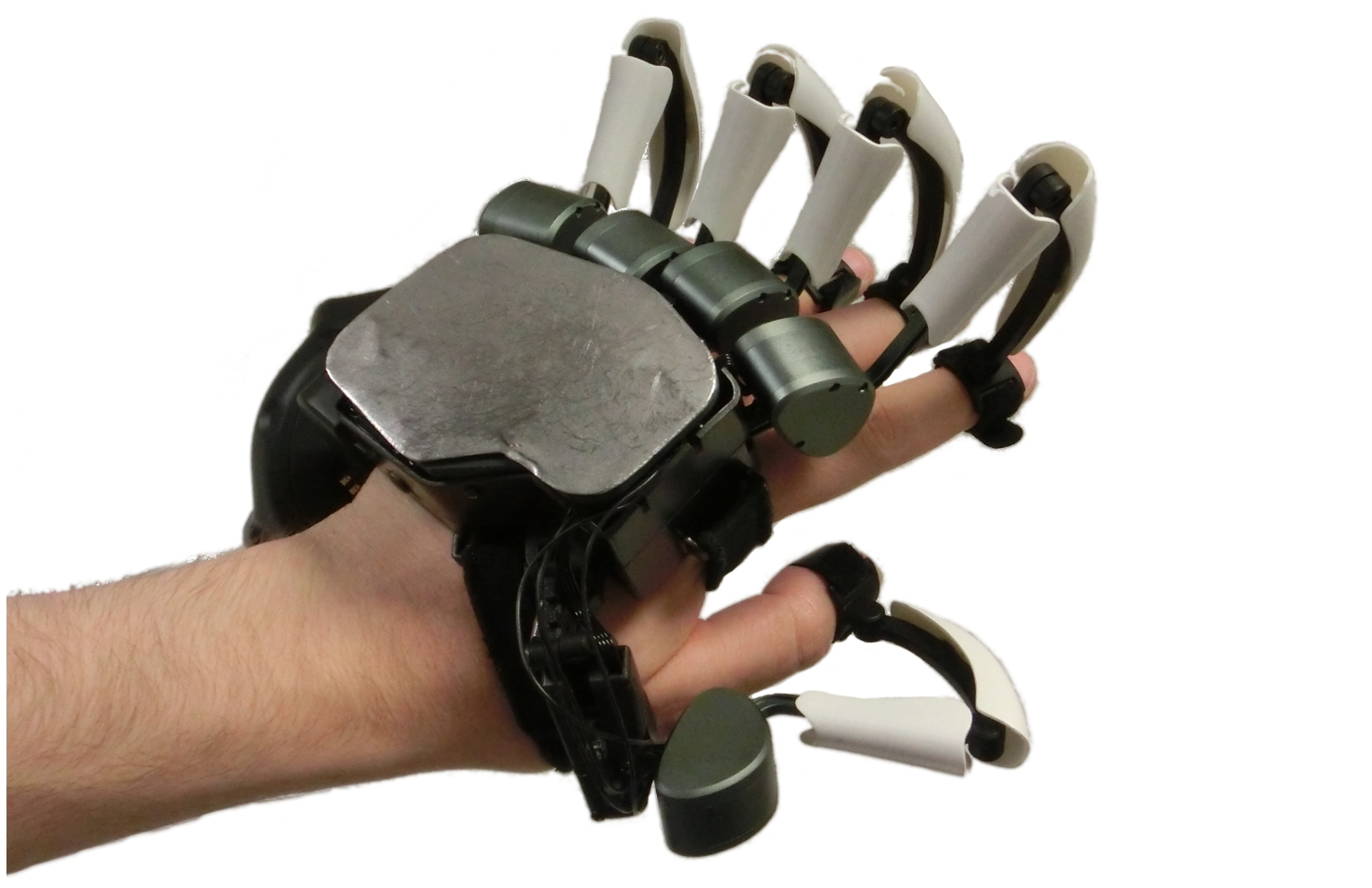}
  \includegraphics[width=0.45\textwidth]{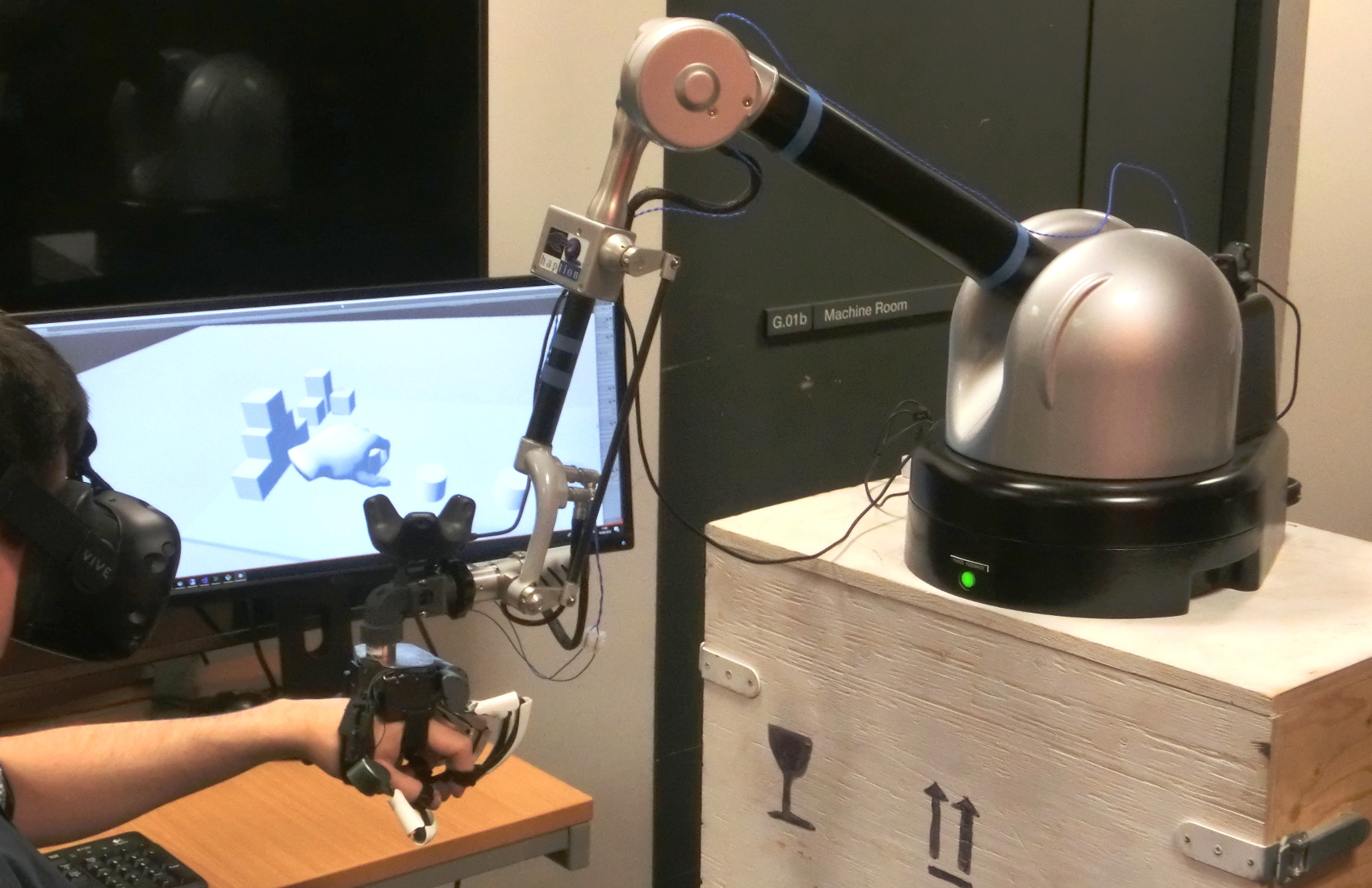}
  \caption{In our prototype implementation of docking haptics, the user wears a hand exoskeleton (Dexmo) with an attached
mild steel plate (Left). Right: As they approach an area where the simulation desires to create the impression of weight, the
exoskeleton is dynamically attached to a grounded six degree-of-freedom robot, a Haption Virtuose 6D, that applies forces
through the plate to the hand exoskeleton.}
  \label{fig:teaser}
\end{figure*}

A grounded device can simulate weight easily, but its range is limited by its mechanical configuration. We suggest that in this situation, the user could be wearing a hand exoskeleton continuously, and this exoskeleton would dock with a grounded device only when necessary to generate ground-referenced force. Figure~\ref{fig:teaser} shows an example of such a situation. Even with a single grounded device, this would alleviate the need for the user to remain with the working volume of the robot.
%when that working volume may not be obviously apparent to the immersed user. 
With multiple robots - either two hand exoskeletons or multiple active or passive grounded devices - the potential design space for novel hybrid devices is very large.

This paper presents the following contributions:

\begin{itemize}
    \item Introduction and initial exploration of the docking haptics concept.
    \item Presentation of a prototype combining a hand exoskeleton and six \ac{dof} grounded force-feedback device.
    \item A preliminary user study demonstrating that the prototype can convey weight sensations to the hand in docked mode.
%    \item An initial exploration of other potential hybrid combinations.
\end{itemize}

\section{Related Work}
\label{sec:related}

Haptic devices can provide both cutaneous and kinaesthetic sensations~\cite{burdea_force_1996,culbertson_haptics:_2018}. In this paper we will focus on the latter, specifically force-feedback devices, but will discuss the generalisation to broader haptic feedback in Section~\ref{sec:discuss}. There are many ways of categorising force feedback devices. 
A common first distinction is between grounded devices and worn devices.
Grounded devices can apply significant forces to the user because they transmit the force through an active or passive device to the ground and so are limited only by the device compliance.
% For example, they can simulate the effect of weight of an object because they can push on, say, the hand of a participant.
In contrast a worn device can transmit forces from one part of the body to another. An example is a hand exoskeleton that can transmit force between fingers and wrist, and thus generate the sensation of grasp restriction. 

Another categorisation is active versus passive. 
Active devices  %purposefully?
change their shape, either to produce force directly (e.g. through a stylus) or to provide different surfaces to encounter. 
Passive devices do not have to be rigid, but are purely reactive, whether held in the hand or encountered.

A final categorisation that we will use is encountered versus encumbering haptics. With the former, the user does not hold or wear a device or object; they reach out to encounter it. In an encumbered system, the user wears or holds devices or objects. The docking haptic concept can utilise devices across all three of these categorisations.

\subsection{Worn and Held Devices}
\label{sec:related:worn}

The use of hand-mounted devices to provide force feedback in \ac{vr} has a long history. The Rutgers Master II was used in \ac{vr} in the mid 1990s~\cite{langrana_integration_1995}. The commercial CyberGrasp system has been available since the early 2000s~\cite{cyberglove_systems_llc_cybergrasp_nodate}. 
The complexity of articulation of the hand has led to a wide variety of devices and actuation methods.
The Rutgers Master II is typical of palm-mounted devices that restrict finger closure with pressure applied on the front of the finger; the CyberGrasp is typical of exoskeletons with actuators and levers mounted dorsally. The application of similar devices in rehabilitation has engendered extensive research on building such devices (see recent reviews~\cite{heo_current_2012,bos_structured_2016}). There has been a resurgence of interest in such devices within \ac{hci} using new actuation techniques. Recent examples include the RML glove~\cite{ma_rml_2015}, Wolverine~\cite{choi_wolverine:_2016} and Dexmo~\cite{Gu2016}. The last of these became a commercial product, and is the glove used in our prototype described in Section~\ref{sec:proto}. Other types of exoskeleton can be used in \ac{vr}. A full arm exoskeleton was developed by Bergamasco et al.~\cite{bergamasco_arm_1994}. A recent review of these devices can be found in Gopura et al.~\cite{gopura_developments_2016}. Our concept could extend to haptics that dock at other points on exoskeletons as discussed in Section~\ref{sec:discuss:other}.

An alternative to hand mounting is to have the user carry a device that changes shape or applies other effects. There is a strong relation to the general area of shape-changing devices (see later in  Section~\ref{sec:related:encountered}), however we focus on devices that are not meant to be seen by the user and so are suitable for use in \ac{vr}. The Shifty device uses a moving weight to change the perception of the shape being held~\cite{zenner_shifty:_2017}. Grabity, generates kinaesthetic cues through grasp restriction and asymmetric skin deformation~\cite{choi_grabity:_2017}. 
NormalTouch and TextureTouch are both hand-held devices that simulate contact normal and coarse texture, respectively~\cite{benko_normaltouch_2016}. 
The Haptic Revolver~\cite{whitmire_haptic_2018} generates shear, texture and shape cues through a wheel that can revolve under the user's finger. 
Each of these could in principle be docked with other devices as we describe later. 

One alternative to our approach is presented by Heo et al.~\cite{heo_thors_2018}. They generate apparently ground-referenced forces in a non ground-fixed manner using large fans. The device has an effectively unrestricted working range but the forces generated are limited. Another approach is the Wireality system which uses wires attached between points on the user's hand and shoulder~\cite{fang_wireality_2020}. Since this is portable, it can work over wide areas, but it can only restricts movement in certain directions and in particular can't support correct gravitational cues.

\subsection{Grounded Devices}
\label{sec:related:grounded}

The archetypal grounded device is the Phantom~\cite{massie_phantom_1994} and its variants: the user holds a stylus and forces can be transmitted from the base. 
However the active range of this, and similar devices, is limited by the mechanical linkage. The Phantom is small and portable, but with limited range and stiffness.
Similar systems vary in their trade-offs. 
The Hapticmaster~\cite{lammertse_hapticmaster_2002} is stiffer and exerts very high forces (100 N), while the Haption Virtuose 6D~\cite{haption_virtuose_nodate} is equally stiff, not as strong, but has more \acp{dof}.

Pairs of grounded devices can be extended to support two hands \cite{talvas_survey_2014}. They can also be extended to support force feedback on the fingers~\cite{galiana_multi-finger_2015}. For example, the Hiro III robot comprises a robot arm with a five-fingered hand attached~\cite{endo_five-fingered_2009}. Each finger is actuated. The user puts the tips of their fingers in small thimbles close to the tips of the robot hand's fingers. An alternative approach can be found with the previously mentioned CyberGrasp hand exoskeleton. Attached to the CyberForce controller~\cite{cyberglove_systems_llc_cyberforce_nodate}, it becomes a grounded device. Two such devices were used together with a HMD to form a seated haptic workstation~\cite{ott_two-handed_2010}. This configuration of grounded device combined with a hand-exoskeleton is similar to one configuration of our hybrid system: when the parts are docked together, our prototype described in Section~\ref{sec:proto} has similar user affordances.

The main limitation of grounded devices for general use in \ac{vr} is their working volume. Several systems have attempted to build larger workspaces by increasing the reach or number of articulations of the robot \cite{ueberle_vishard10_2004, gosselin_large_2008, gosselin_widening_2007}. The Haption Inca system \cite{haption_inca_nodate} is a novel type of cable-actuated large-scale workspace, based on the Spidar concept~\cite{sato_spidar_2002}. This mechanism could be implemented at a variety of scales. Alternatively a high precision controller can be connected to a wider area moving platform as in the Haption Scale 1 system \cite{haption_scale1_nodate}. Finally, the high precision controller could be connected to a mobile platform~\cite{peer_new_2008,pavlik_interacting_2013}. Our concept of docking haptics extends such systems to be more general: if a mobile platform is available, it might be recruited into an ensemble system.

\subsection{Encountered Devices}
\label{sec:related:encountered}

The devices covered so far require the user to hold or wear a device. Encountered devices allow the user to reach out and touch surfaces with their bare hands. Encountered type devices may be passive or active. In a passive system, the real world is represented within \ac{vr} so that virtual objects may have some solidity~\cite{insko_passive_2001}. At the very simplest, an object that the user can pick up and hold can be considered a type of passive or encountered device~\cite{hinckley_passive_1994}. Otherwise, the passive object might be furniture or walls that the user might engage with in various ways.  There has also been a lot of interest in active encountered displays that can change shape or be reconfigured~\cite{mcneely_robotic_1993}. Within the HCI community there has been enormous interest in the past few years in shape-changing interfaces (see reviews~\cite{coelho_shape-changing_2011,sturdee_analysis_2018}). We propose that both active and passive encountered devices might be recruited into docking systems.

\section{Docking Haptics Concept}
\label{sec:concept}

The docking haptics concept proposes that by dynamically docking haptic devices we can explore a much larger design space for haptic feedback. By docking, we mean one device attaching to another with a temporary joint. The joint may be rigid, or have some \acp{dof}.

While some types of multi-part shape-changing robot fall under this definition (e.g.~\cite{zhao_robotic_2017}) we are interested in exploring the space by combining the best capabilities of different classes of device. 

\begin{figure*}[!t]
\centering
  \subfloat[][]{%
    \includegraphics[width=\figtwoscale\textwidth]{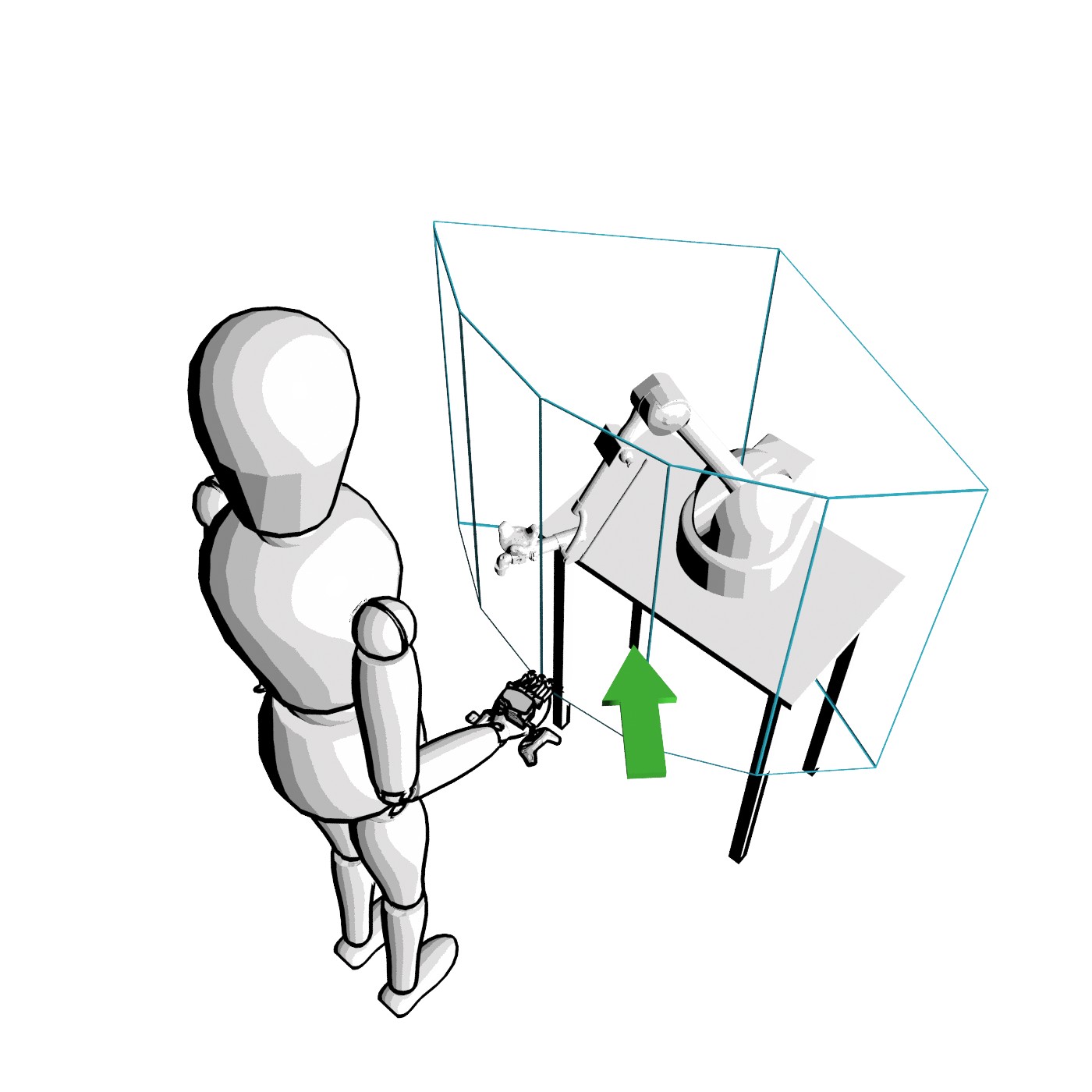}%
  }
  \hspace{0.3cm}
  \subfloat[][]{%
    \includegraphics[width=0\figtwoscale\textwidth]{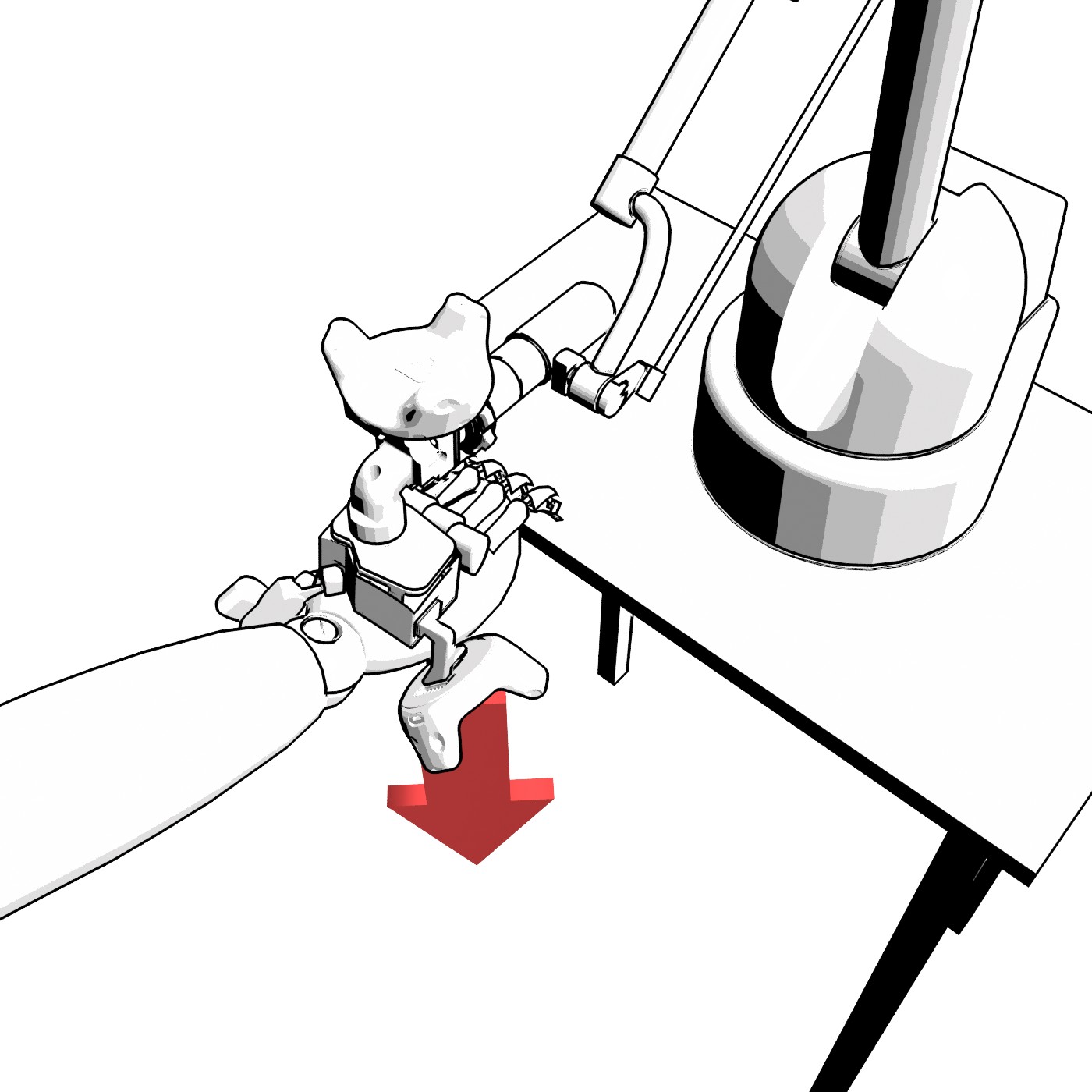}%
  } 
  \hspace{0.3cm}
  \subfloat[][]{%
    \includegraphics[width=\figtwoscale\textwidth]{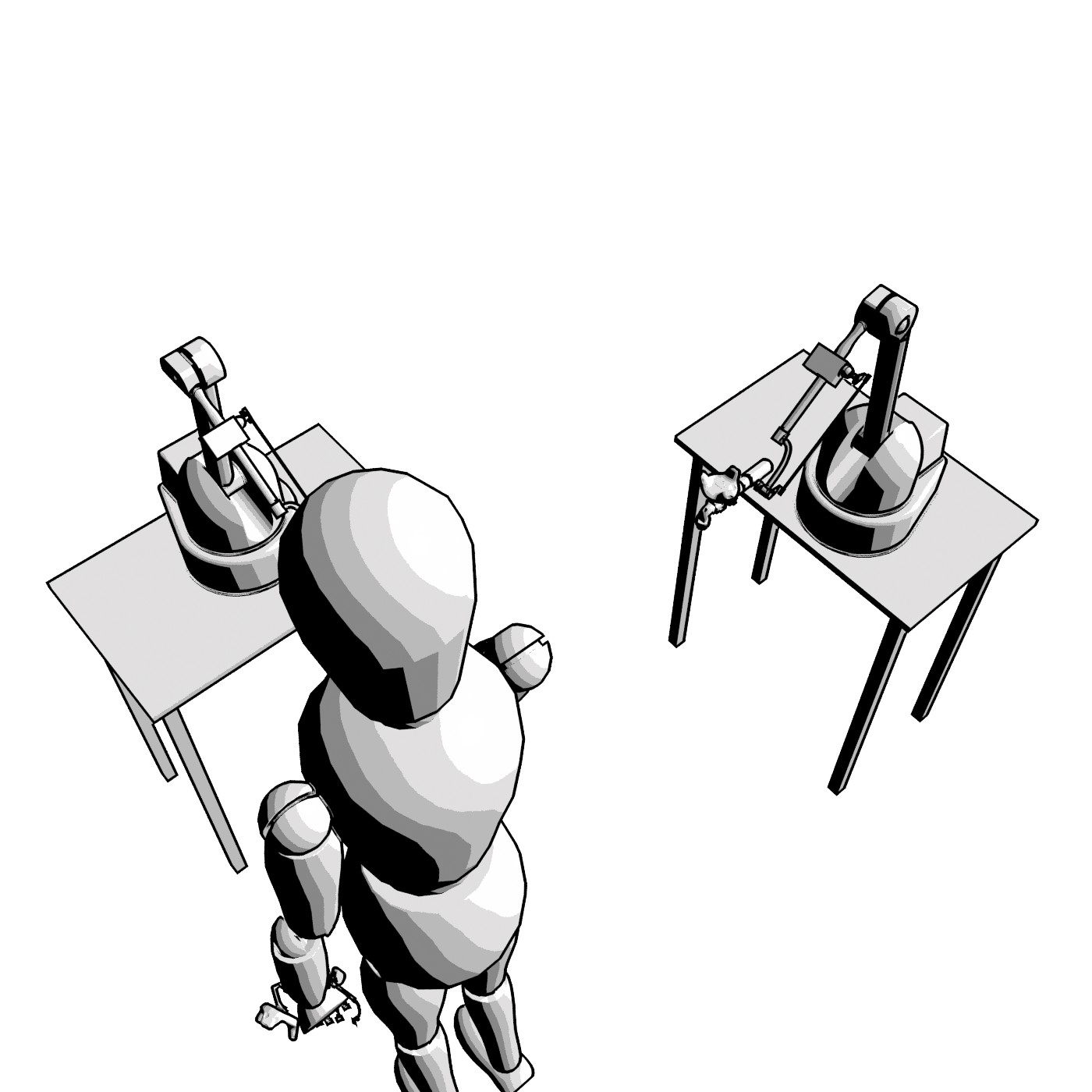}%
  }
  
  \subfloat[][]{%
    \includegraphics[width=\figtwoscale\textwidth]{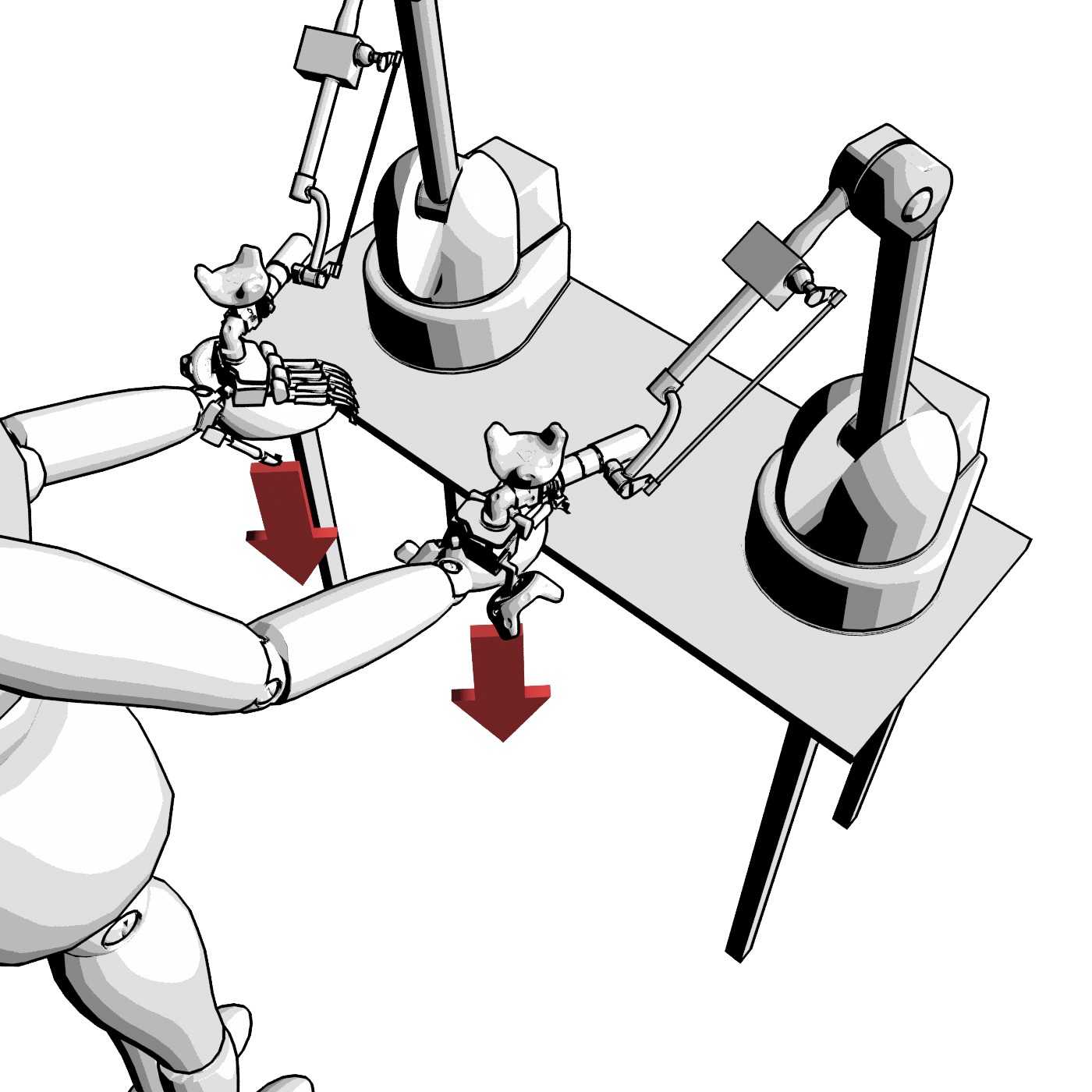}%
  }
  \hspace{0.3cm}
  \subfloat[][]{%
    \includegraphics[width=\figtwoscale\textwidth]{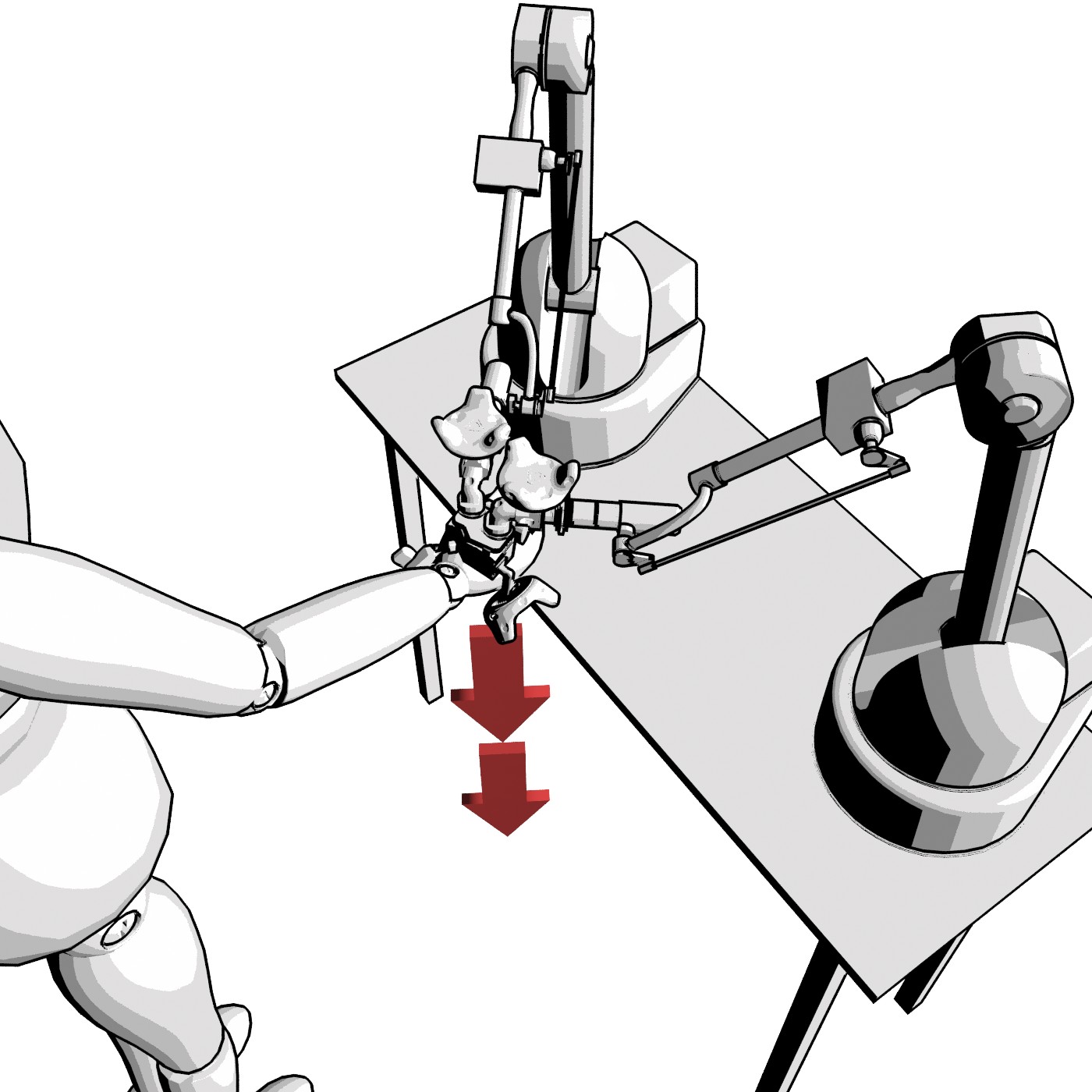}%
  }
  \hspace{0.3cm}
  \subfloat[][]{%
    \includegraphics[width=\figtwoscale\textwidth]{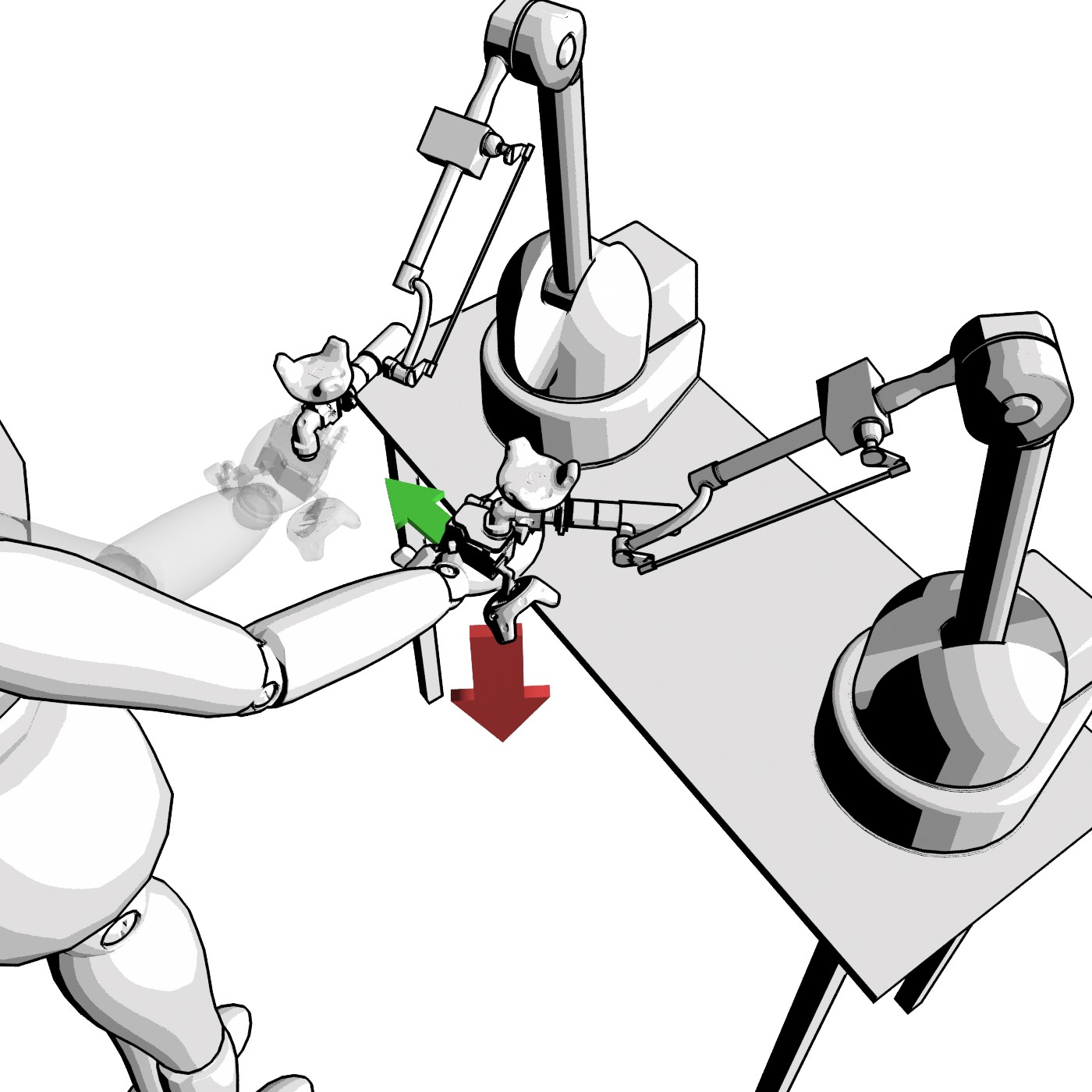}%
  }
  
  \subfloat[][]{%
    \includegraphics[width=\figtwoscale\textwidth]{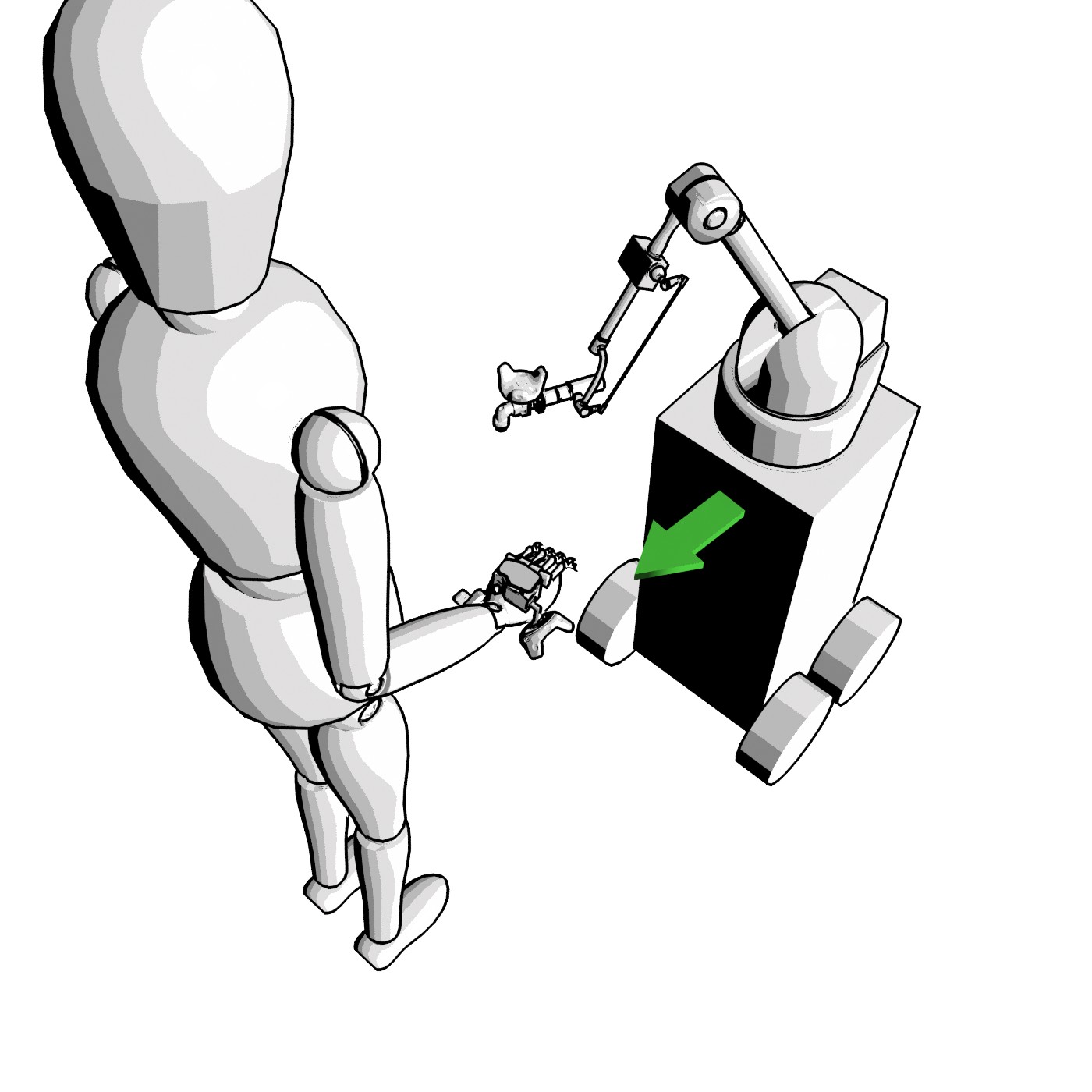}%
  }
  \hspace{0.3cm}
  \subfloat[][]{%
    \includegraphics[width=\figtwoscale\textwidth]{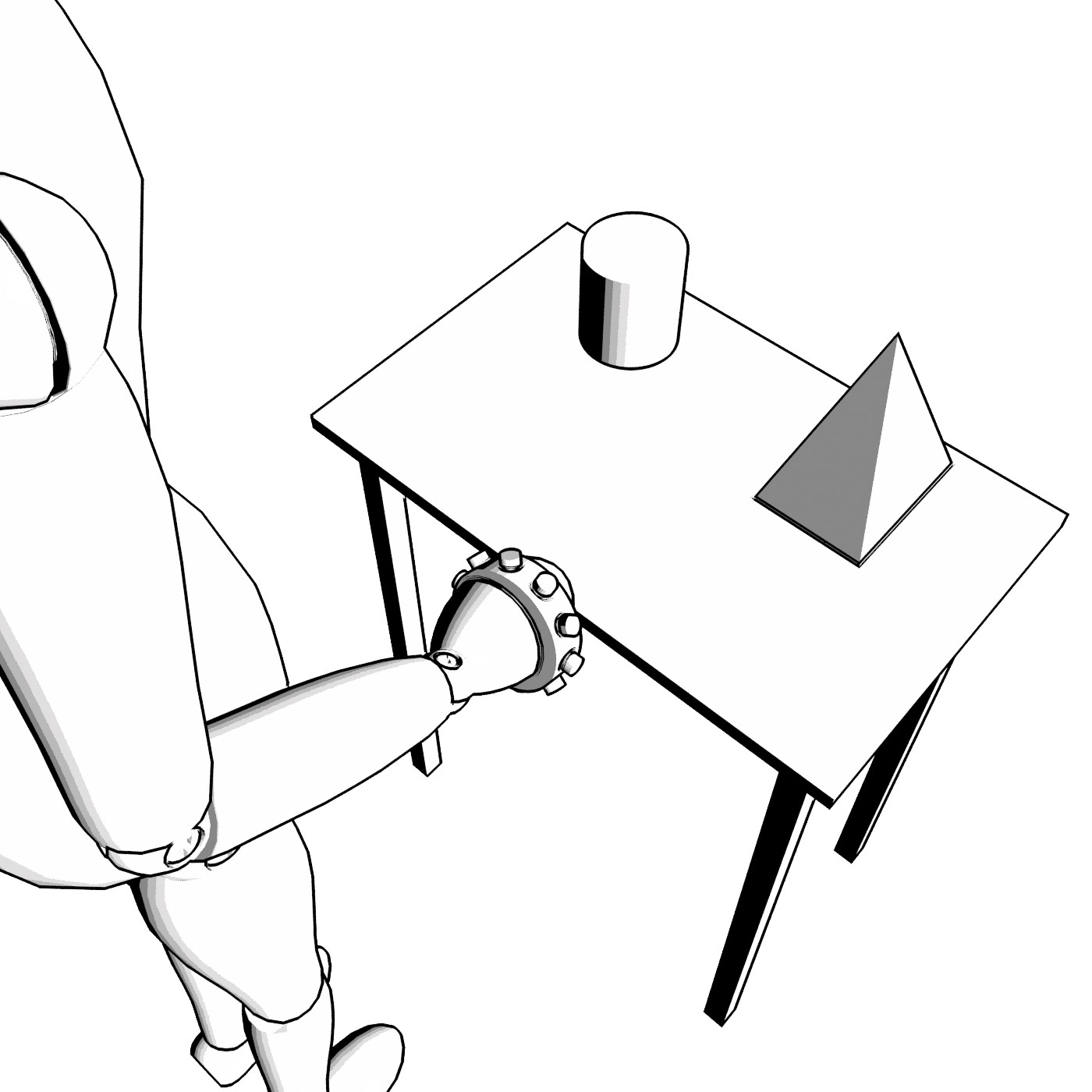}%
  }
  \hspace{0.3cm}
  \subfloat[][]{%
    \includegraphics[width=\figtwoscale\textwidth]{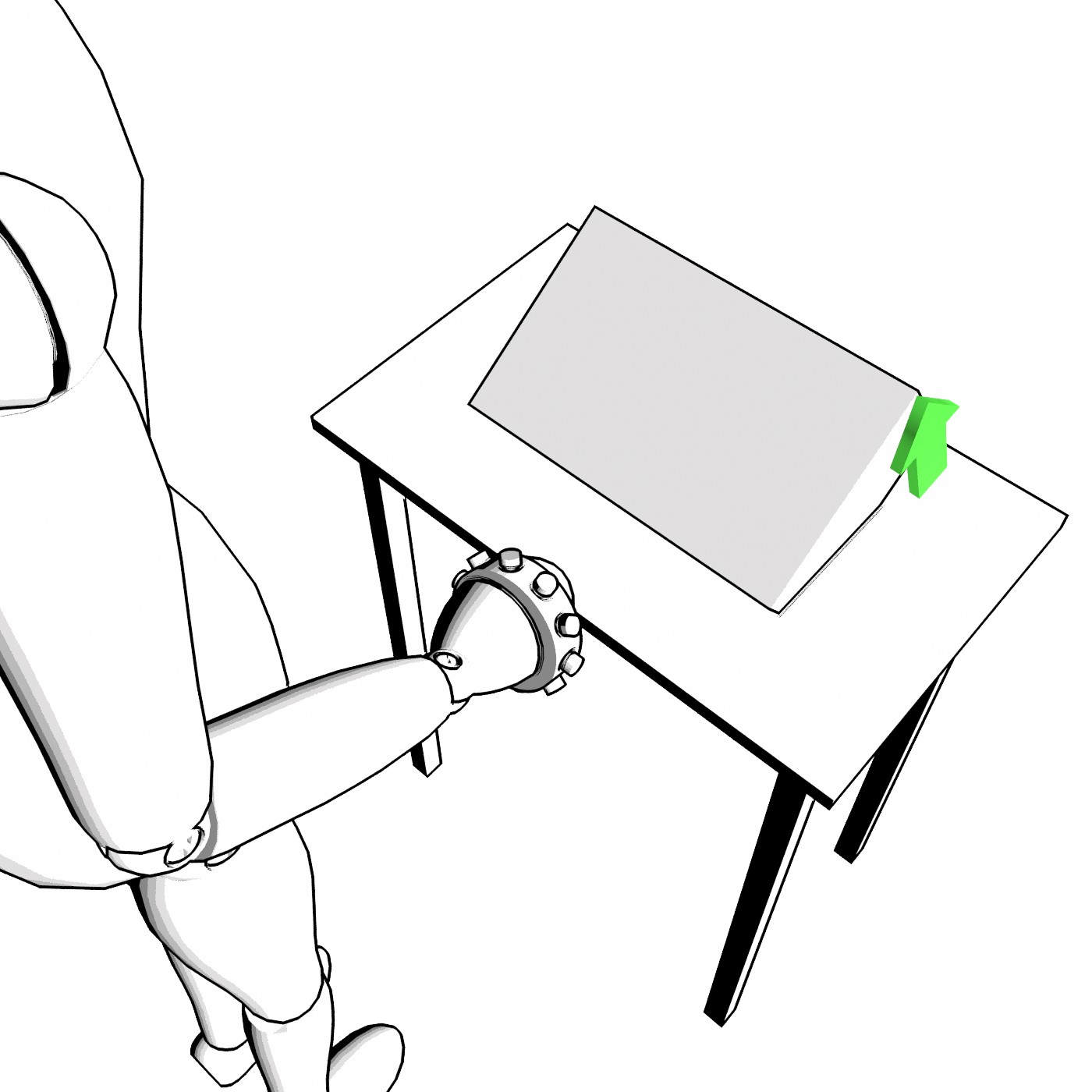}%
  }
  
  %  \stackunder{
  %\includegraphics[width=0.32\textwidth]{test_Camera001.jpg}
  %}{(a)}
  % \stackunder{
  %\includegraphics[width=0.32\textwidth]{test_Camera002.jpg}
  %}{(b)}
  %\stackunder{
  %\includegraphics[width=0.32\textwidth]{test_Camera003.jpg}
  %}{(c)}
  % \stackunder{
  %\includegraphics[width=0.32\textwidth]{test_Camera004.jpg}
  %}{(d)}
  % \stackunder{
  %\includegraphics[width=0.32\textwidth]{test_Camera005.jpg}
  %\stackunder{
  %\includegraphics[width=0.32\textwidth]{test_Camera006.jpg}
  %}{(f)}
  %\stackunder{
 %\includegraphics[width=0.32\textwidth]{test_Camera007.jpg}
 %}%{(g)}
 %\stackunder{
 %\includegraphics[width=0.32\textwidth]{test_Camera008.jpg}
 %}{(h)}
 % \stackunder{
 %\includegraphics[width=0.32\textwidth]{test_Camera009.jpg}
 %}{(i)}
  \caption{Example configurations and operations of docking haptics. (a) The user moves (green arrow) their hand towards the grounded robot working volume (blue cage) the robot moves to intercept their hand. (b) Once docked the grounded device can apply net force (red arrow) to the hand. (c) Supporting the hand in two working volumes. (d) Two-handed operation. (e) Two robots apply double the net force (red arrow). (f) As the hand moves (green arrow) out of the working volume of the first robot, the second grounded robot takes over the application of the net force. Other scenarios include: mobile grounded robot interaction (g), docking a hand-held controller with passive haptic objects (h), or docking a hand-held controller with a shape-changing device (i).}
  \label{fig:cartoons}
\end{figure*}

\subsection{Examples}
\label{sec:concept:eg}

A first example, which we have prototyped as described in Section~\ref{sec:proto}, is the combination of a hand exoskeleton and grounded mechanical arm (see Figure~\ref{fig:teaser} for an image of the prototype). As illustrated in Figure~\ref{fig:cartoons}(a)-(b), as the user moves their hand into the target area of the grounded device, the grounded device can intercept and attach to the hand. The two robots then act as one device that could be modelled by a single kinematic chain. Note that in Figure~\ref{fig:cartoons}, we use the equipment in our implementation to illustrate the concept, % (c.f. Figure~\ref{fig:mount}), 
but other configurations and equipment are possible. 

When docked, the combined robot has some combination of the degrees of freedom of the two underlying robots. Table \ref{tab:examples} shows some examples. Connecting a device  to a grounded robot explicitly constrains it to that robot, and the type of constraint will depend on the joint (see Section \ref{sec:concept:types}). Please note that the description of the Dexmo glove is somewhat simplified as it has different articulations for each powered joint, see \cite{Gu2016}.

\begin{table*}[]
    \centering
    \tiny
    \caption{Docking two devices (Device A and Device B) creates a hybrid device in a number of ways. Our prototype combines a Virtuose 6D and a Dexmo glove}
\begin{tabular}{l|lllll}
Device        &  Type &  Translation Volume & Rotational Volume &  Translation Force & Rotational Force  \\ \hline
Device A      &   Grounded             & $X_A \times  Y_A \times Z_A$                      &  $RX_A \times  RY_A \times RZ_A$               & $3 \times N_A $       & $3 \times Nm_A $  \\ 
Device B      & Hand-worn              & $\infty$                                          &  6 of $R_B$                                    & -                     & 6 of  $Nm_B $ \\
Figure 2a/2b  &  A with B              & $X_A \times  Y_A \times Z_A$ or $\infty^1$        &  $RX_A \times  RY_A \times RZ_A$               & $N_A$                 & $3 \times Nm_A$ + 6 of $Nm_B $ \\
Figure 2e     & 2 of A with B          & $X_A \times  Y_A \times Z_A$ or $\infty^2$        &  $RX_A \times  RY_A \times RZ_A $ + 6 of $R_B$ & $2N_A + 2 \times N_A$ & $3 \times 2Nm_A$ + 6 of  $Nm_B $ \\
Figure 2f     & 2 of A with B          & $2X_A \times  Y_A \times Z_A$ or $\infty^2$       &  $RX_A \times  RY_A \times RZ_A $ + 6 of $R_B$ & $3 \times N_A$        &  $3 \times Nm_A$ + 6 of $Nm_B $ \\
Virtuose 6D   & Grounded               & $1330mm \times 575mm \times 1020mm$               &  $330^o \times 130^o \times 270^o$             & $3 \times 9.5 Nm$     &  $3 \times 1Nm $ \\
Dexmo         & Hand-worn              & $\infty$                                          &  5 of $165^{o5}$                                       & -                     &  5 of $0.5Nm^5$\\
Our Prototype & Virtuose 6D with Dexmo & $1330mm \times 575mm \times 1020mm$ or $\infty^4$ &  $330^o \times 130^o \times \infty^4$ + 6 of XX& $3 \times 9.5 Nm$     &  $ 2 \times 1Nm^4$ + 5 of $0.5Nm$\\
\multicolumn{6}{l}{ } \\
\multicolumn{3}{l}{\textsuperscript{1} Acts as B outside range of A} & \multicolumn{3}{l}{\textsuperscript{3} Acts as Dexmo outside of range of Virtuose 6D} \\

\multicolumn{3}{l}{\textsuperscript{2} Acts as B outside range of both A} & \multicolumn{3}{l}{\textsuperscript{4} In the prototype we lose one degree of rotational constraint around the joint} \\
\multicolumn{3}{l}{\textsuperscript{6} Coincident with the metacarpal of each finger}\\ \multicolumn{6}{l}{} % todo: more accurate word than coincident...

\end{tabular}

    \label{tab:examples}

\end{table*}

By using multiple grounded devices we expand the range of possibilities. We could provide grounded feedback at multiple locations: this could fit with the simulation of a workspace with, for example, tool selection on one bench or rack and a workpiece on another (Figure~\ref{fig:cartoons}(c)). Thus, some haptic feedback can be provided over the entire space, with enhanced feedback in important task-related areas. Alternatively, two grounded devices could be adjacent: this could support two-handed interaction; double the force potential on a single hand; or a larger working space for one-hand by handing over between the two grounded devices (Figure~\ref{fig:cartoons}(d)-(f)).

The robots could be of different types: the grounded device might be mobile or an encounter-style robot; the user might hold a simple hand-held device with very simple haptic feedback that could dock with passive haptics or a shape-changing device (Figure~\ref{fig:cartoons}(g)-(i)).

\subsection{Types of Docking}
\label{sec:concept:types}

Docking creates a connection between two bodies. If we consider this as a mechanical problem, we would be forming {\em kinematic pairs}~\cite{hartenberg1964kinematic}. Examples of such pairs include revolute or hinged joints, prismatic joints and planar joints. However, we want to effect the joint temporarily, so we could consider mechanisms such as electromagnets, hydraulics (e.g. suction) or mechanical linkages (e.g. a grabber).

\begin{figure*}[!t]
  \centering
  \subfloat[][]{
  \includegraphics[width=\figthreescale\textwidth]{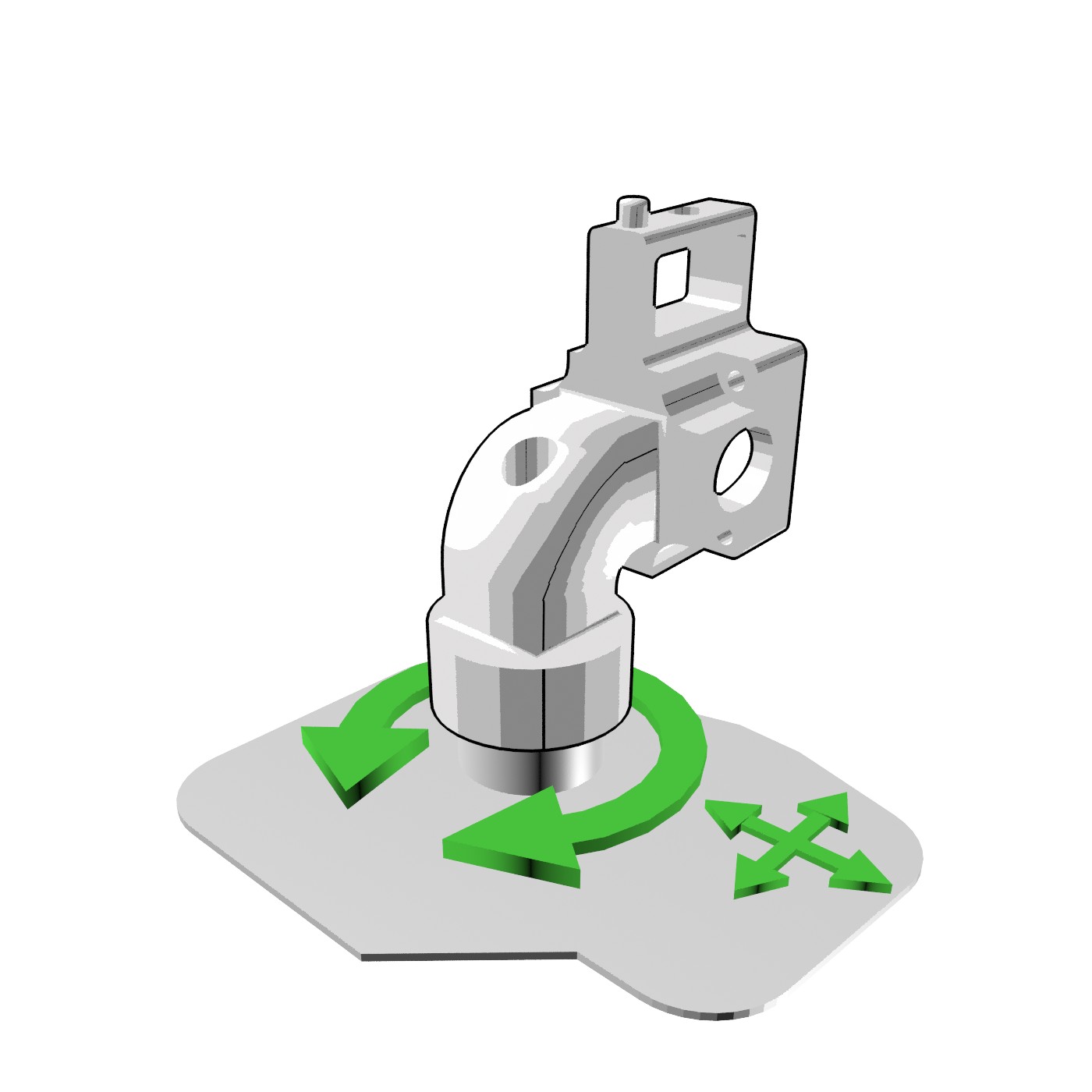}
  }
  \hspace{0.3cm}
  \subfloat[][]{
  \includegraphics[width=\figthreescale\textwidth]{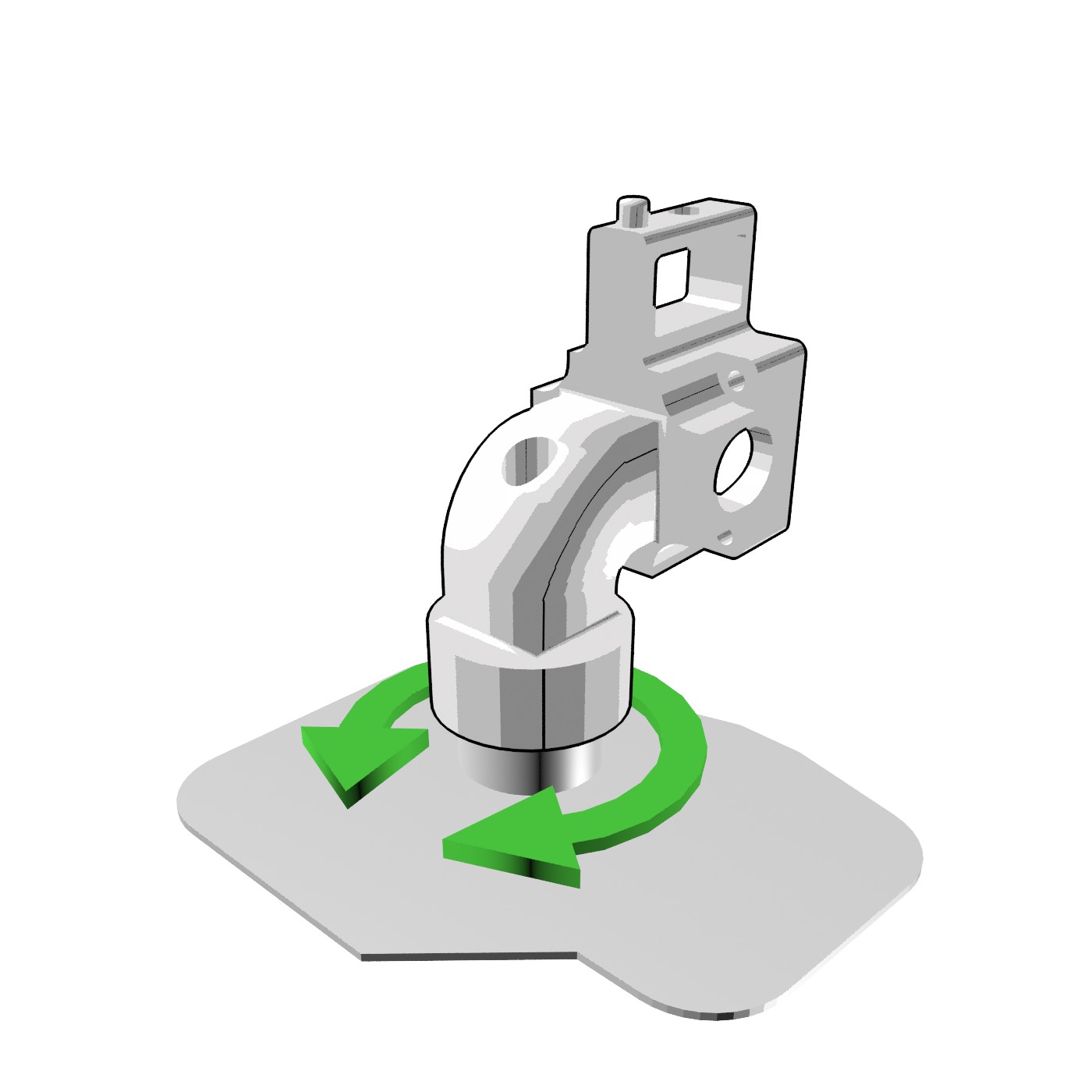}
  }
  \hspace{0.3cm}
   \subfloat[][]{
  \includegraphics[width=\figthreescale\textwidth]{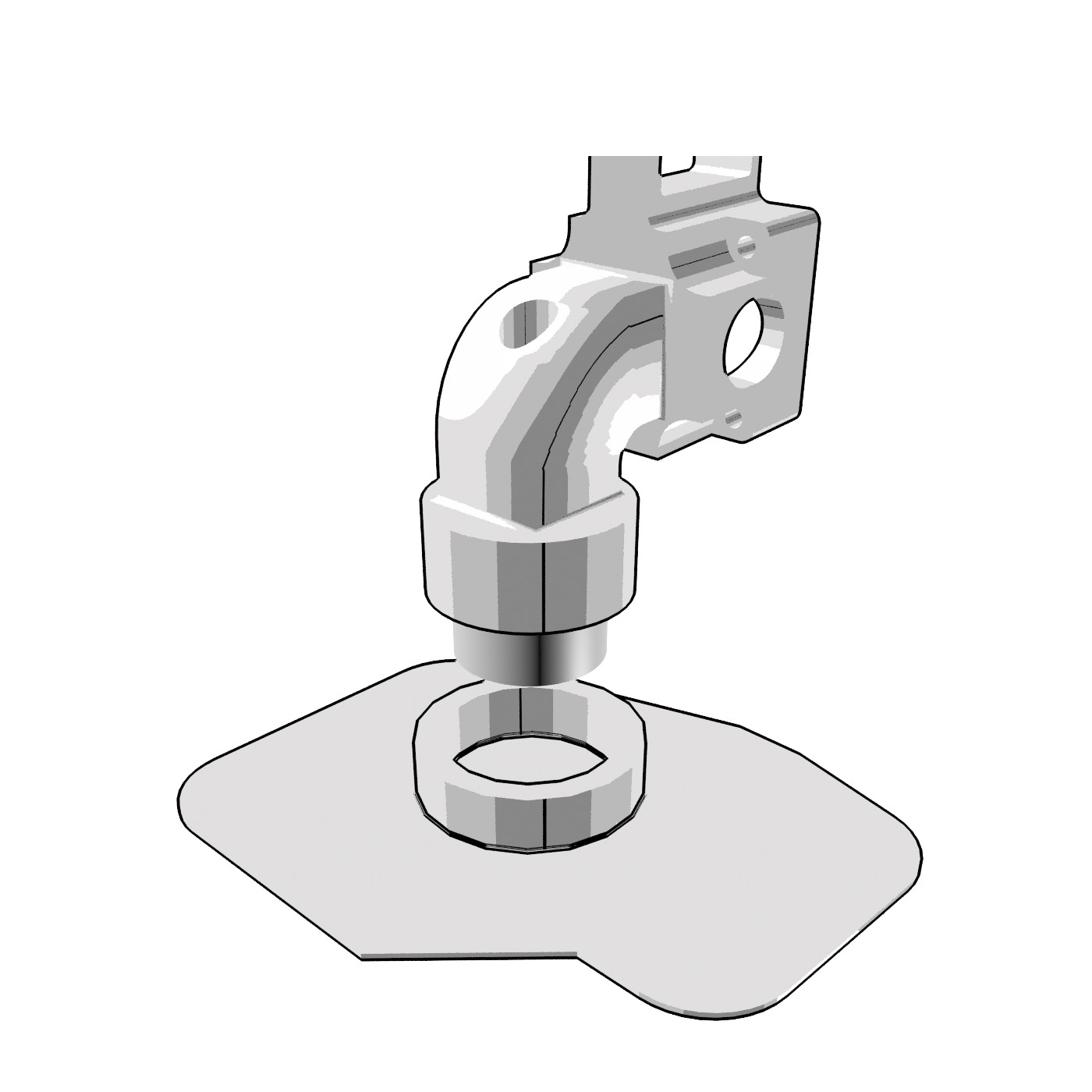}
  }

  \subfloat[][]{
  \includegraphics[width=\figthreescale\textwidth]{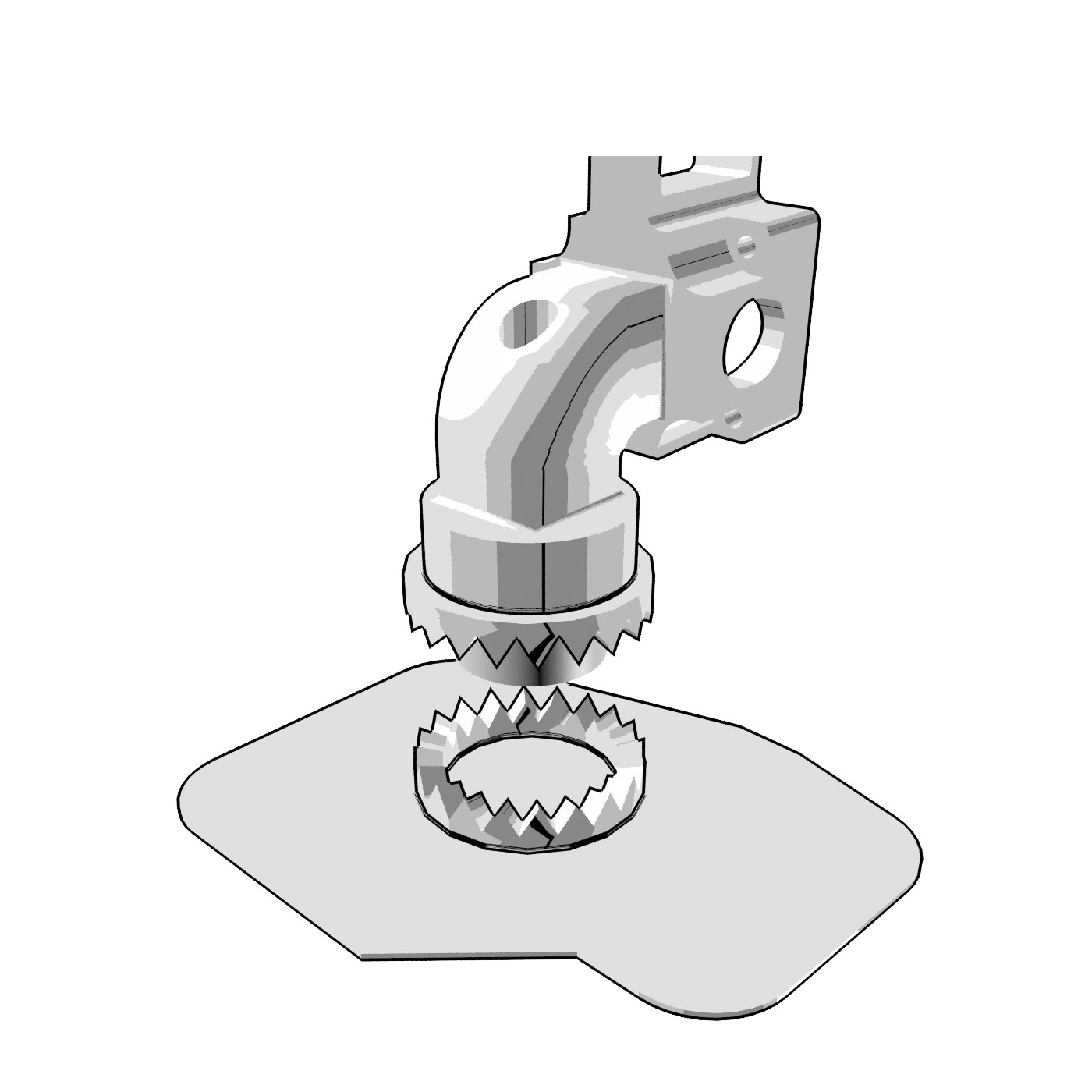}
  }
  \hspace{0.3cm}
  \subfloat[][]{
  \includegraphics[width=\figthreescale\textwidth]{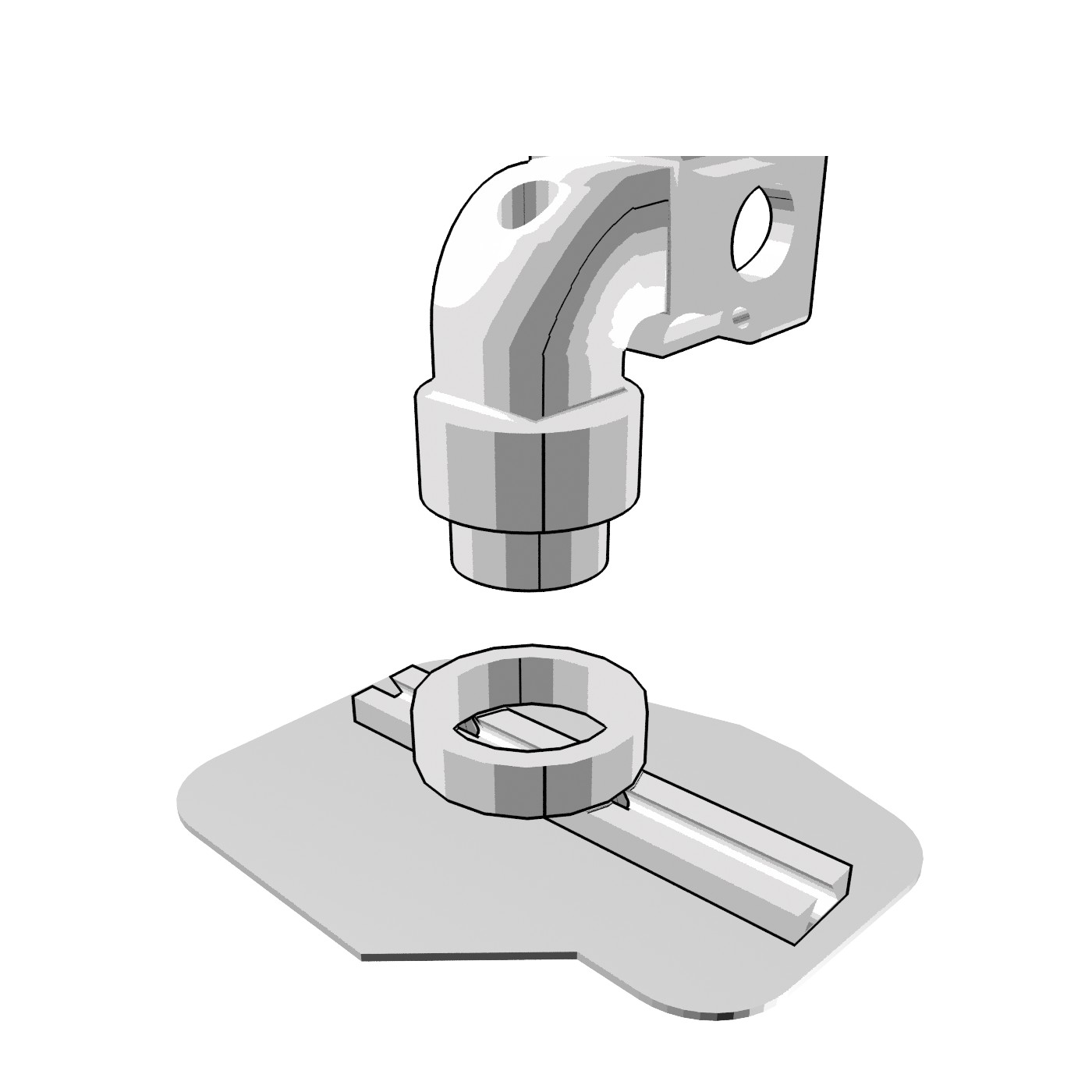}
  }
  \hspace{0.3cm}
   \subfloat[][]{
  \includegraphics[width=\figthreescale\textwidth]{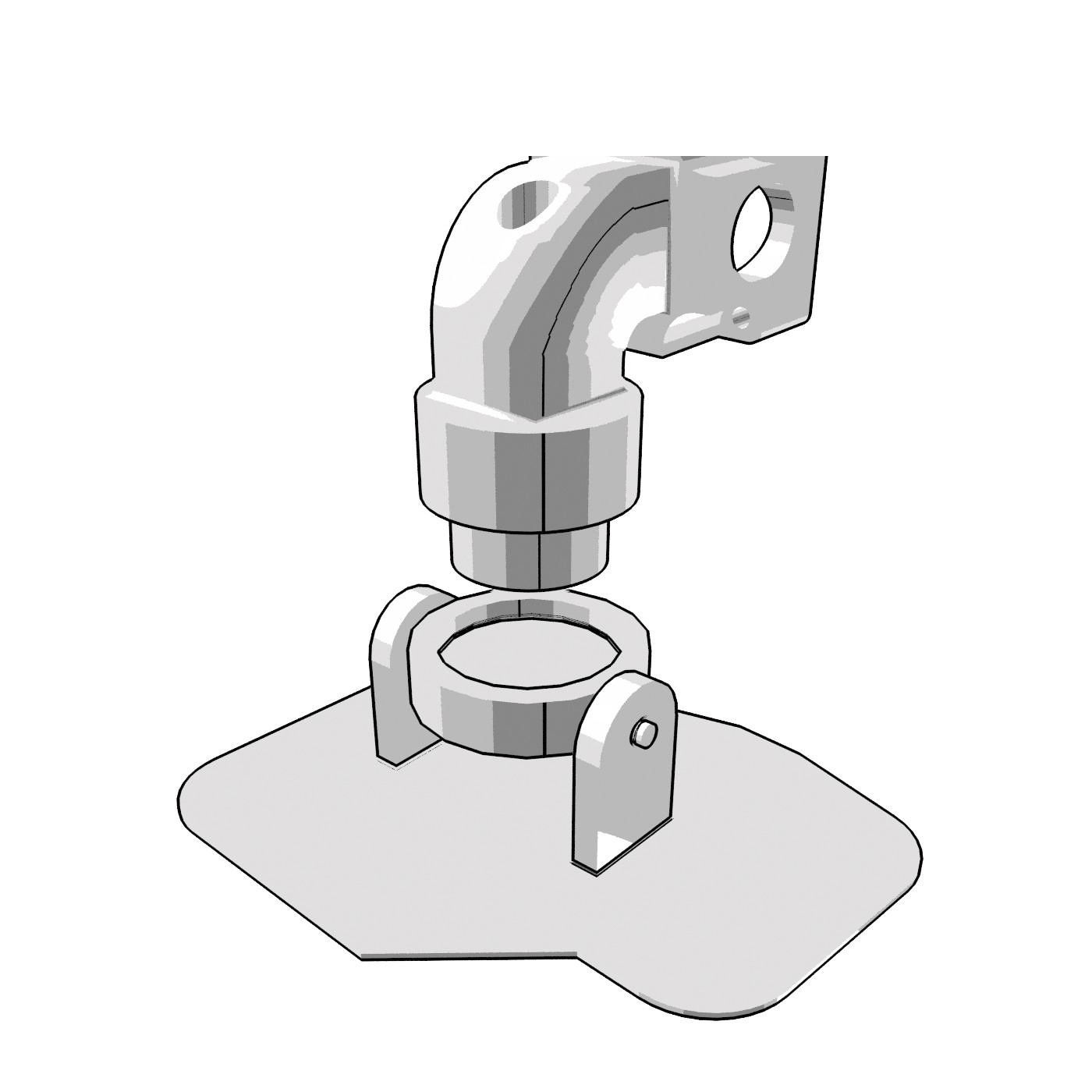}
  }

  \caption{Example joints. (a) With a low magnetic strength, the joint will slip and turn on the plate. (b) With a stronger force, the joint will not slip, and will turn with some resistance. (c) If the magnet targets a holder on the surface, the joint can be made precisely and turning more freely with lower power to the magnet. (d) Teeth on the magnet holder and docking surface prevent rotation. (e) A sliding joint can be made. (f) Other degrees of freedom can be relaxed. }
  \label{fig:joints}
\end{figure*}

Electromagnetic coupling in an attractive option. We can create surfaces that can attach at many points by using ferromagnetic materials on one device and magnetic coils on the other. We can also design around the breaking force of the joint: we can set the magnetic strength so that the devices decouple  without software intervention before enough force is exerted to endanger the user or machine.
A simple attachment of a magnet to a plate, as discussed in Section 4, depends on the friction between the plate and the magnet once the electromagnet is activated. This could act as a plate joint (two translation and one rotation around plane normal) or hinged joint (one rotation around plane normal), see Figure~\ref{fig:joints}(a)-(b). 
If we can target a small plate in a holder, we can ensure the axis of rotation (Figure~\ref{fig:joints}(c)) or we can couple the devices using a profile around the joint so that they are effectively rigid under lower magnetic force (Figure~\ref{fig:joints}(d)). That small plate could itself be on a joint (e.g a slider or hinge, Figure~\ref{fig:joints}(e)-(f)), or other combinations.

Joint type is important to the range of forces that can be applied. We may lose \acp{dof}, such as only being able to apply force tangential to the attachment plane, or in the case of our prototype, lose the ability to apply torque to the hand. 
This suggests a trade-off between joint simplicity and force transfer capability.
We may need to model the joint's behaviour so it can be properly considered as part of the kinematic chain. This suggests some implementation options such as targeting very precise docking, or allowing opportunistic docking where we measure the relative transforms of the two bodies afterwards to allow proper kinematic treatment. Further, each body could have one or more attachment surfaces, so the docking could be effected with whichever pairs of surfaces provided the best range for the resulting kinematic chain and the task required. 

So far, we have considered joints that have well-defined constraints and free directions. In future work, more complex interactions such as electromagnetic force at a distance, elastic or flexible connections, etc., could be considered where the mechanism itself contributes unique forces to the haptic rendering.

\section{Prototype Implementation}
\label{sec:proto}

We constructed a proof-of-concept dockable haptic workspace. The wearable was a Dexmo glove and the arm was a Haption Virtuose 6D with an electromagnet-based docking mechanism. Both units were controlled by the same Unity application. The complete apparatus is shown in Figure~\ref{fig:teaser}:Right. 

\subsection{Dexmo Glove}
The Dexmo glove by Dexta Robotics (Figure~\ref{fig:teaser}:Left, with our modifications) is a new iteration of a passive-admittance based device presented by Gu et al. \cite{Gu2016}. The current version has 5 force-feedback \acp{dof}, 11 sensed \acp{dof}, and 10 uninstrumented \acp{dof}. Force feedback is provided by motors that pull normal to the fingertips via a link-bar arrangement.
The Dexmo glove has an admittance based API. Local feedback loops with contact-drum~\cite{Springer2002} like behaviour are parameterised by normalised target rotations and unit-less spring constants for each actuator. The domain of the normalised parameters are the rotations at the extremes of finger flex and extension, measured during a calibration for each user.

As the Dexmo is under-instrumented, the forward model that drives the virtual hand must make some assumptions. Namely, that the user adopts a power-grasp pose~\cite{Cobos2010}, and the \ac{dip} \& \ac{pip} are linearly proportional to the \ac{mcp} \cite{Li2007a,Hrabia2013,Cobos2008}. Finger abduction (spread) is measured directly.
The virtual hand consisted of a graphical model from the Dexmo SDK and a skeleton that included joint limits. Bio-mechanical models of the hand are available \cite{VanDerHulst2012,Buchholz1992,Cobos2008}, but for this project we calibrated ours by eye as the skeleton itself was not physically correct.
% (the flex & abduction joints co-indcident, whereas in real life they are a saddle joint. the lengths are also not right)
%
The normalised parameters from the Dexmo sensors directly interpolated joint rotations between their limits. Their limits represent the extreme hand poses that users adopt during calibration. The Dexmo is tracked using an HTC Vive tracking puck.

\subsection{Haption Arm}

The Virtuose 6D by Haption is a robotic arm designed for ground-fixed tool-based haptics. It has 6~\ac{dof}, able to set both tool position and orientation within its workspace. 
%The Virtuose comes with a detachable hand tool, which we replaced with our mount, see Figure~\ref{fig:mount}. 
The Virtuose supports both impedance and admittance-based control. We use it in admittance mode. The Virtuose SDK works with end-effector or virtual tool transforms in a Cartesian coordinate system.

The programming model for admittance mode is to set target transforms and speeds in a high-frequency (1kHz) callback. These are used to compute the parameters for the robot's control loops. The user-API communicates with a process that runs the robot over UDP, so the SDK can be used on any computer. Indeed, the arm was connected to a different machine than the HMD.

\subsection{Magnetic Dock}

The robot attached to the hand using an electromagnet. This was a generic 12v electromagnet with a peak force of 5kg and diameter of 25mm. The Dexmo cover was replaced with a mild-steel plate. The Virtuose tool was replaced with an attachment hosting the magnet and a tracker for open-loop docking (Section~\ref{sec:docking}). The magnet power supply was controlled with simple serial commands to a microcontroller over USB. We did not measure the current drawn by the magnet in use, but the magnet would become warm to the touch.

The magnet exerts high forces axially, but it is easy to pivot off. For the prototype, we fixed the tool orientation and instructed users to keep their hands flat to avoid applying any torque to the attachment. Relaxing these constraints is discussed in Section~\ref{sec:mounts}.

The working volumes and force characteristics of the Dexmo and Virtuose devices and the hybrid device resulting from docking with the specific magnetic joint, are shown in the lower part of Table \ref{tab:examples}.

\subsection{Docking}
\label{sec:docking}

\def\Effect{\mathit{Effect}}
\def\EffectorToTool{\mathit{EffectorToTool}}
\def\ToolToEffector{\mathit{ToolToEffector}}

\begin{figure}[t]
    \centering
    \includegraphics[width=0.5\linewidth]{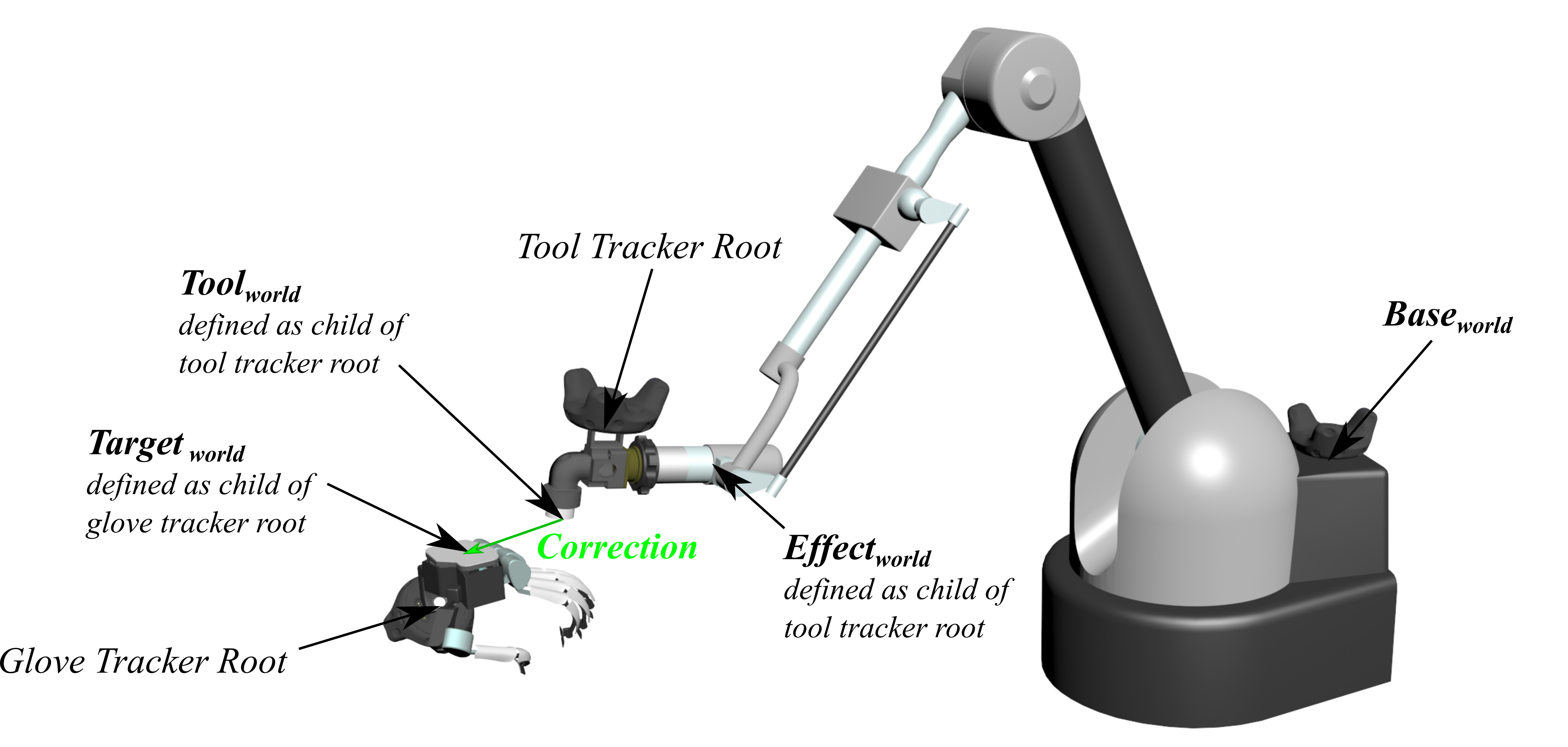}
    \caption{Coordinate systems for docking}
    \label{fig:transforms}
\end{figure}

Tracking and interception is a well studied problem in robotics, in the forms of navigation guidance and visual servoing \cite{Lin1989,Agah2004,Mehrandezh2000,Borg2002,Chwa2005,Kunwar2006}. For this prototype we simply attempted to match the transform of the magnet to a target point on the hand (Pure Pursuit).
Control of the Virtuose for docking and interception was performed in the arm's base frame, the native frame of the admittance API, with Unity's coordinate system.
The Virtuose API takes a target transform (displacement and orientation) of the end-effector pivot. To compute this, we transform the target by the inverse of the magnet's local transform. This gives the true effector target.
The Virtuose maintains an estimate of its current position, however the accuracy we found was too low for reliable interception. 
Instead we use another Vive puck to track the effector and compute corrections, as per Equation~\ref{eqn:tracking}, where $\Effect'_{forward}$ and $\Effect_{forward}$ are respectively the target and current effector transforms to and from the API. World space transforms are determined by the external trackers, which operate in a coordinate system referenced to the real world by the SteamVR calibration performed outside the application. Refer to Figure~\ref{fig:transforms}. 

\begin{comment}
\begin{flalign}
\begin{aligned}
&   Effect_{local} = Base^{-1}_{world} * Effect_{world}                         &\\        
&   Target_{local} = Base^{-1}_{world} * Target_{world}                         &\\
&   EffectorToTool = Effect_{local}^{-1} * (Base^{-1}_{world} * Tool_{world})   &\\
&   ToolToEffector = EffectorToTool^{-1}                                        &\\
&   Effect'_{local} = Target_{local} * ToolToEffector                           &\\
&   Correction = Effect_{local}^{-1} * Effect'_{local}                          &\\
&   Effect'_{forward} = Effect_{forward} * Correction                                    
\end{aligned}
\label{eqn:tracking}
\end{flalign}
\end{comment}

\begin{flalign}
\begin{aligned}
\Effect_{local} & = Base^{-1}_{world} * \Effect_{world}      &\\        
Target_{local} & = Base^{-1}_{world} * Target_{world}        &\\
\EffectorToTool & = \Effect_{local}^{-1} * (Base^{-1}_{world} * Tool_{world})   
&\\
\ToolToEffector & = \EffectorToTool^{-1}                     &\\
\Effect'_{local} & = Target_{local} * \ToolToEffector        &\\
Correction & = \Effect_{local}^{-1} * \Effect'_{local}       &\\
\Effect'_{forward} & = \Effect_{forward} * Correction                                    
\end{aligned}
\label{eqn:tracking}
\end{flalign}

\subsection{Virtual Environment}

The \ac{ve} was constructed in Unity 2018 and used the inbuilt PhysX engine to perform the simulation. The Dexmo glove provides a managed library that was integrated directly with Unity. The Virtuose C library was thinly wrapped with P/Invoke. The Dexmo should not be updated at more than 30Hz while the Virtuose must be updated at no less than 30Hz. The Dexmo and magnet controller ran in the main Unity thread, while the Virtuose SDK ran its own thread to issue callbacks, and in these callbacks we implemented force control and tracking \& interception, based on parameters set from the main thread.

\subsection{Physics Simulation}

PhysX is an impulse-based rigid-body simulation. The graphical hand model was provided with kinematic colliders. That is, PhysX would apply impulses to dynamic objects to resolve collisions with the visual geometry of the hand.
We drove the simulation using rigid body colliders as opposed to force-reflection in order to allow users to grasp items. The power-grasp assumption and limited instrumentation of the glove make it very difficult for users to correctly balance point-forces on rigid bodies for stable grasping, or facilitate force-closure. With colliders, the simulation incorporates additional forces that cannot be actuated by the glove, reducing transparency but increasing intuitiveness and controlability. Dynamic objects are left to the physics engine to simulate. We observe the impulses that drive their behaviour and use them to control the haptic devices.

\subsection{Robot Control}

The Dexmo glove has an admittance-based API. The force feedback parameter is an angular position in the same space as the sensor parameters - effectively an interpolant between the extremes of flex and extension. 
Once set the glove will prevent the user exceeding the specified rotation/finger flex.
%, representing a pose that the user should not exceed. 
To compute this, we detect collisions between the distal phalange colliders and world geometry. We let the physics engine resolve the collision by moving the world, but before this happens we compute a hypothetical de-penetration transform that moves the phalange colliders instead. From these hypothetical positions, we determine numerically their extent between the extreme poses, and thus the `contact-drum' parameter. The stiffness of the surface was fixed.

We use the Virtuose in admittance mode, though for force-feedback we implement an impedance based controller on top of this, as it was most natural to work with forces at this stage. The controller computes an appropriate displacement each frame based on Hooke's Law, the spring-constant being defined by the stiffness of the robot's control loop.

The forces on the hand are rendered by the local loop on the glove itself. The forces for the arm are computed from the impulses applied by the hand-colliders. These are transformed into forces and summed. The result is transformed into the arm's local space and applied through the effector.
This is because when a user is squeezing an object, equal and opposite forces will be applied to opposing hand colliders. 
When the object is supported externally, world referenced forces will be transferred through world geometry, and hand-referenced forces will cancel.
If the hand is supporting an object against gravity, inertia or other geometry, world referenced forces will be transferred through a subset of hand colliders. The net forces in this case are non-zero. The palm-fixed actuators cannot drive these, so they are applied through the arm.

\section{Proof of Concept Experiment}
\label{sec:exp}

To prove the docking haptics concept we performed a within-subjects study. 18 people from within UCL performed weight sorting tasks in \ac{vr} under the three conditions in Table~\ref{tab:conditions}: \emph{Free} where there was feedback on the Dexmo Haptic only, \emph{Docked} where the Dexmo attached to Virtuose arm but the Virtuose was not actuated and  \emph{Force Feedback} when the two robots are attached and the Virtuose was providing force feedback. The principal purpose was to establish the difference in participant experience between these three conditions with the expectation that the participants would find it easier to sort the weights in the Force Feedback condition. This was assessed through an objective measure of their success in the sorting task and also through a subjective rating. Section \ref{sec:questionnaire} includes other questions about the subjective experience. The study was performed under UCL Research Ethics Committee approval 4547/012. Participants were compensated \textsterling 10 for taking part. 

\begin{figure}[t]
    \centering
    \includegraphics[width=0.6\linewidth]{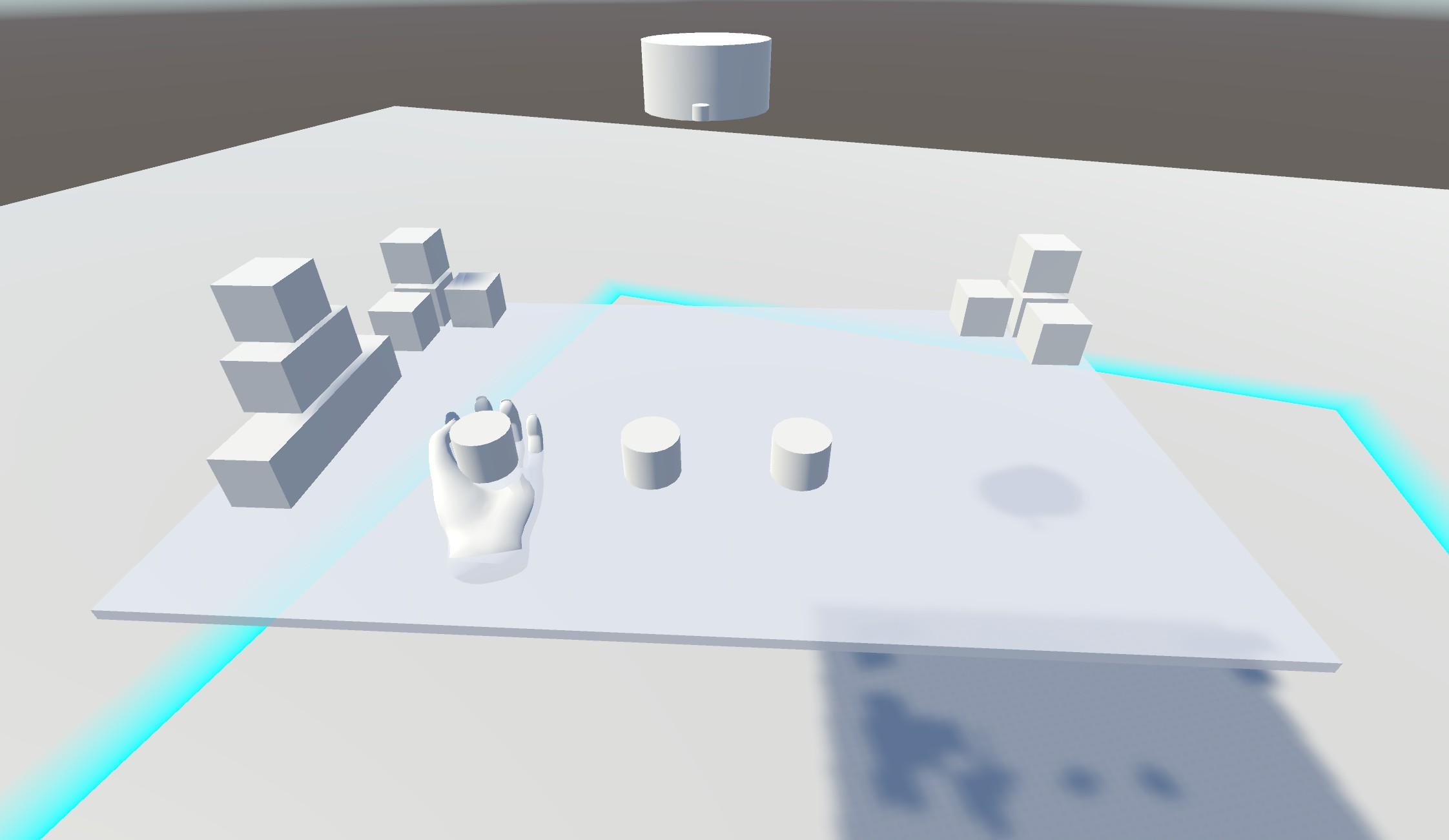}
    \caption{Participant's view of the virtual world}
    \label{fig:experimentscreenshot}
\end{figure}

\begin{table}[t]
    \caption{Experiment conditions}
    \small
    \centering
    \begin{tabular}{l|ccc}
     \multirow{2}[0]{*}{Feature} & \multicolumn{3}{c}{Condition} \\
          & \multicolumn{1}{l}{Free} & \multicolumn{1}{l}{Docked} & Force Feedback \\
          \hline
    Dexmo Haptic Feedback & \multicolumn{1}{c}{\checkmark} & \multicolumn{1}{c}{\checkmark} & \checkmark \\
    \hline
    Robot Attached &       & \multicolumn{1}{c}{\checkmark} & \checkmark \\
    \hline
    Robot Force Rendering &       &       & \checkmark \\
    \hline
    \end{tabular}%
    \label{tab:conditions}%
\end{table}%

\subsection{Apparatus \& World}

The robotic apparatus is the system described previously. The VR headset was an HTC Vive. 
An example of participants' view of the world is shown in Figure~\ref{fig:experimentscreenshot}. Participants could pass their virtual hand through the translucent worktop, but the other items were physically simulated and provided haptic feedback. The three cans had their rotations constrained. Other items did not have a role in the task, but could be manipulated by the participant during the induction stage as part of their explorations.

\subsection{Task \& Procedure}

Utility was evaluated via an objective sorting task in which participants were presented with three visually identical cans and asked to arrange them in order of increasing weight using ``whatever cues they could perceive'' to distinguish between them.

Participants were inducted with verbal and written instructions. After giving verbal consent, they were assisted in putting on the Dexmo Glove and a calibration procedure was performed outside of \ac{vr}. The participants then put on the headset and positioned themselves in front of the virtual desk by moving in the real world.
They were then given an unlimited period in which to interact with world (conditions identical to Free) in order to become comfortable in the virtual space and with the hand's interaction with it.

Once the participants were ready, the trials were completed in immediate succession. The experimenter informed them which condition was next, reset the can positions, and where necessary docked the arm. Participants had an unlimited amount of time to arrange the cans, after which the procedure repeated for the remaining conditions. After all conditions, participants exited \ac{vr} and completed a questionnaire.

The condition order was randomized per-participant. Can weights were always $0.01$, $0.15$ and $0.3$. The order was randomized per trial. These masses were tuned to be perceptibly different while not generating destabilisingly large impulses. The masses are nominally in kg, but the robots' actuations were not physically accurate.

Participants had to use their right hand only for the task as only the right-hand Dexmo glove was modified. We did not check which hand was the participant's dominant hand, but none expressed a problem completing the tasks with their right hand. The experiment took roughly 30 minutes, with 15 minutes being immersed in the VR system.

\subsection{Questionnaire}
\label{sec:questionnaire}

To evaluate the quality of experience, participants were given a questionnaire. Participants were given the same five questions about each condition and told to judge the experience against the real world. Responses were on a five-point (negative (1) to positive (5)) linear scale.
Q1: How easy was it to pick up an object? Q2: How well were you able to explore using touch? Q3: How easy was it to tell the difference between the weights? Q4: How well could you disambiguate surface texture? Q5: How well were you able to move as you desired?

The rationale for these questions was that we would expect that there would be a difference on Q3 between  Force Feedback and the other two. We did not have a prior hypothesis about Q1 as the specific act of picking is supported in the same way in each condition in that the participant grabs the object and feels restriction through the Dexmo glove. We would also expect that the limited range of the Virtuose would mean that in both conditions with docking, participants would rate their movement as being restricted (Q5). Q4 was a control question as the system did not render or simulate the dynamics of surface texture.

\subsection{Results}

\begin{figure}
    \centering
    \includegraphics[width=0.6\columnwidth]{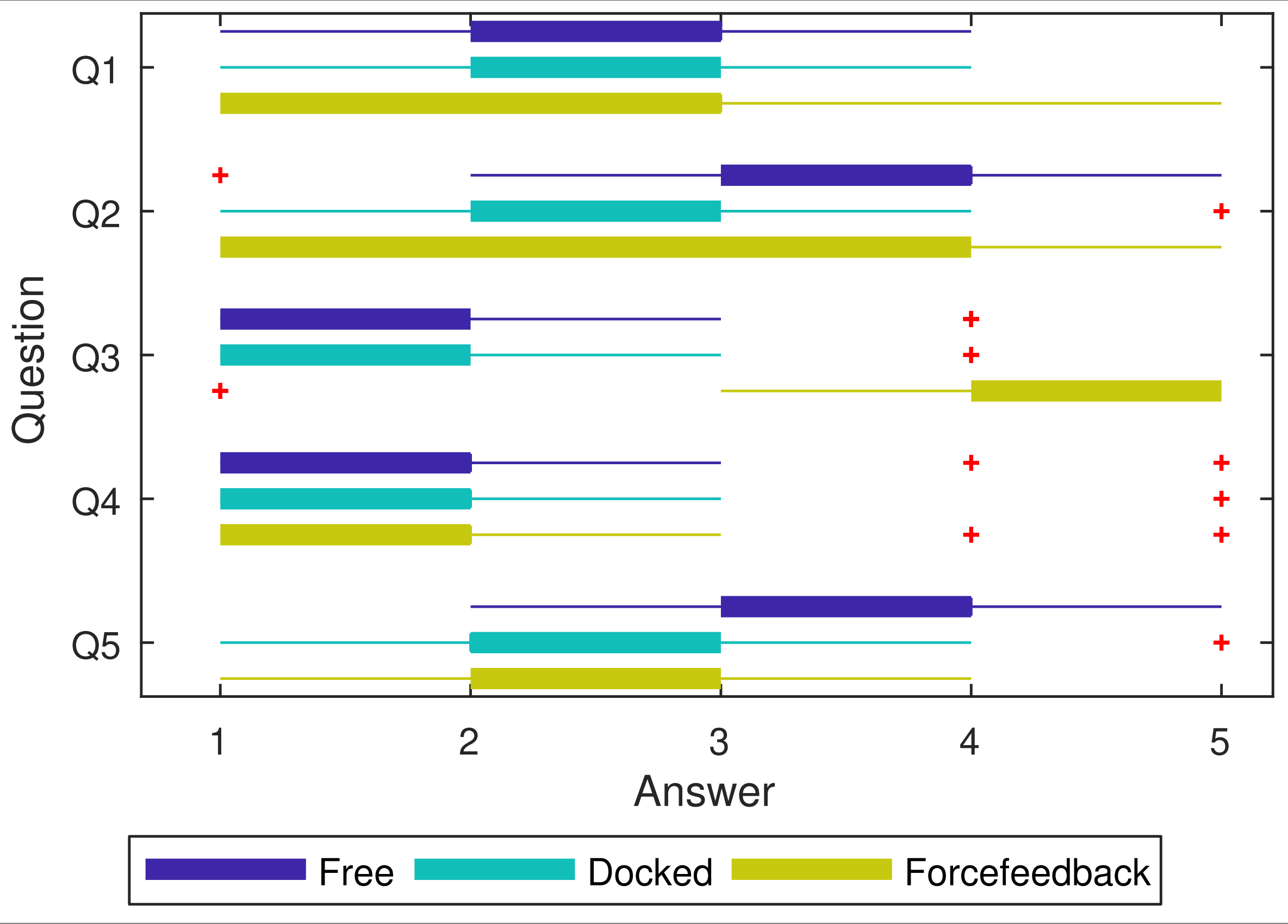}
    \caption{Box plots of questionnaire responses}
    \label{fig:questionnaireresponses}
\end{figure}

\begin{table}[htbp]
  \centering
          \caption{Mean and standard deviation for questionnaire responses}

    \footnotesize
    \begin{tabular}{l|l|l|l|l|l}
          & Q1    & Q2    & Q3    & Q4    & Q5 \\
        \hline
    Free  & 2.83/0.76 & 3.35/1.01 & 1.88/1.09 & 1.94/1.12 & 3.77/0.97 \\
        \hline
    Docked & 2.44/1.01 & 2.55/1.16 & 1.83/1.01 & 1.88/0.99 & 2.50/1.11 \\
        \hline
    Forcefeedback & 2.55/1.21 & 2.50/1.25 & 4.05/1.26 & 2.00/1.10 & 2.44/1.01 \\
    \end{tabular}%
  \label{tab:means}%
\end{table}%

\begin{table}[htbp]
    \caption{p-values for significant pair-wise differences between Conditions for each question}

    \centering
    \small
    \begin{tabular}{l|r|r|r|r|r}
    \multirow{2}[0]{*}{Condition Pair} & \multicolumn{5}{c}{Question} \\
    %\cline{2-6}
        & \multicolumn{1}{l|}{Q1} & \multicolumn{1}{l|}{Q2} & \multicolumn{1}{l|}{Q3} & \multicolumn{1}{l|}{Q4} & \multicolumn{1}{l}{Q5} \\
        \hline
    Free-Docked &       & 0.0078 &       &       & 0.0020 \\
    \hline
    Free-Forcefeedback &       & 0.0039 & 0.0005 &       & 0.0026 \\
    \hline
    Docked-Forcefeedback &       &       & 0.0002 &       &  \\
    \hline
    \end{tabular}%

    \label{tab:questionnaireinteractions}%
\end{table}%

The questionnaire answers for each condition are shown as a box plot in Figure~\ref{fig:questionnaireresponses} with means and standard deviations in Table~\ref{tab:means}. Table~\ref{tab:questionnaireinteractions} shows the condition pairs that received significantly different responses according to the Wilcoxon signed rank test.

A one-way ANOVA on the number of correctly ordered pairs per trial ($F_{2,51}=7.42,p=0.001$) showed that user performance was significantly better ($\overline{83\%}$) in the Forcefeeback condition than in the other two, statistically indistinguishable, conditions ($\overline{53\%}$). The condition had no significant effect on completion time.

As can be seen, users find the docking mechanism inhibits their range of motion but does not significantly affect their ability to pick up objects, though in general they did not find it easy to grasp objects. Some participants commented after the experiment that they found the experience tiring, especially with the robot attached.  As expected users gained no benefit from docking itself, and only in the Forcefeedback condition were they able to distinguish between the weights, as indicated by both questionnaire results and task performance. The responses in the Forcefeedback were more varied than in the other conditions. This is possibly due to varying individual tolerances to the force jitter caused by the suboptimal physics simulation.

\section{Discussion and Future Work}
\label{sec:discuss}

\subsection{Mount Design}
\label{sec:mounts}

Our mount was successful in that it supported interception, docking and the conveyance of weight. Users felt that it limited their ability to explore however.
We suspect this is due to the artificial rotation constraints, imposed because the attachment cannot transfer torque.
One solution would be for the robot to match the orientation of the hand, actively compensating for user-applied torque. %This requires more precise control of the robot than we demonstrate, and more importantly, such an approach would not be able to apply torque.
Another solution is a mechanical linkage such as a universal joint. The magnet could remain attached with optimal alignment without active compensation. The control system would have to model the behaviour of the joint to ensure forces were always applied normal to the attachment. %however, and the arm would still not be able to apply torque.

A mount design could control different \acp{dof} with different techniques. For example, the magnet could enter a tube rather than dock anywhere on a plate. The tube would mechanically limit off-normal rotations, but axial forces remain entirely magnetic. %This design is mechanically simple and highly rigid, but would require very accurate interception by the grounded robot.

\subsection{Robot Capabilities}

Where rigid docks are used, robot range-of-motion will become perhaps the most important characteristic. Active compensation is necessary for any practical system, as the Dexmo glove and robot together are heavy. Active compensation may require more precise control, but it could also be performed with force sensors only, which the Virtuose supports.

The Virtuose had errors in its forward kinematics and as such required an external tracker to intercept the hand.
These error sources present challenges but not necessarily significant ones.
Works on robot guidance often express rendezvous algorithms in terms of relative velocities (e.g. \cite{Mehrandezh2000,Borg2002,Chwa2005,Kunwar2006}). 
With external tracking, all that is required for these is to transform velocity commands so that resulting accelerations are correct in the global frame - for example, with a change-of-basis recomputed each tracking frame. Such an approach is far simpler than attempting to model the error sources across the entire workspace. Since external tracking is required for the base location and wearable location, there is little disadvantage to tracking the tool.

Future systems will have to explore more complex algorithms than we demonstrate, to support faster hand motion, robot grasping and avoidance of obstacles, most notably the hand itself. The specific problem that docking haptics presents is how to anticipate touch so that docking is not delayed, but then how to do this smoothly so that there is no impulse from the attachment itself.

\subsection{Physics Simulation}

Future implementations will need a better integration in order to render physically correct forces.  Driving the simulation with colliders instead of force-reflection resulted in an intuitive and stable simulation even at low (for haptics) frame rates. 
Game-optimised engines may have difficulty with complex arrangements of colliders like those of the hand, resulting in distracting jitter and extreme responses.
More stable simulation approaches such as Position Based methods~\cite{Bender2012}, or advanced simulation engines such as MuJoCo~\cite{todorov_mujoco:_2012}, could provide a better experience.

\subsection{Other Configurations}
\label{sec:discuss:other}

Our prototype demonstrates just one possible docking configuration. Section~\ref{sec:concept} suggested other approaches. One aspect to explore is the type and direction of docking. Even with magnetic docking, we had the option of mounting the magnet on the hand exoskeleton or the grounded robot; swapping this around may provide more opportunities. For example, one obvious configuration is an encounter-style robot that instead of only rigid objects (e.g. \cite{araujo_snake_2016}), holds plates or surfaces that can be docked with.
This suggests that the hand-device, be it a held controller, or a hand exoskeleton, could have multiple magnets to connect.

Another area to explore is the combination of docking with haptic retargeting~\cite{azmandian_haptic_2016}. Given that joints may be stronger in some directions than others, and that users seem relatively insensitive to small relative position offsets, we can perhaps construct docking configurations that appear to generate forces from a wider range of directions and strengths than is physically realisable.

\section{Conclusion}
\label{sec:conclusion}

In docking haptics, a system dynamically couples different components at run-time to build the most appropriate hybrid haptic device for a particular task and environment.
As the user transitions throughout the virtual and physical worlds the capabilities of the hybrid system change. 

We prototyped an example system from a Dexmo exoskeleton glove and  Haption Virtuose arm, to explore a hybrid grounded/un-grounded system. Using pure-pursuit, the arm was able to intercept the glove and exert forces in cooperation with the glove to render weight. Using a straightforward net force calculation on different collider sets we could control the system with minimal knowledge of the individual device dynamics.
In a user study we validated the concept by demonstrating that even with its limitations, our docking haptics prototype conveys weight with sufficient fidelity to significantly improve performance in an object sorting task.

Docking haptics could be used in a variety of applications where it is important to combine freedom of movement with force feedback. For example, following Figure \ref{fig:cartoons}(c) we can imagine applications that provide multiple workbenches (e.g. mechanical training or medical training) that the participant might move between. Following Figure \ref{fig:cartoons}(f) docking haptics might  support applications that might require a larger workspace than provided by the grounded robot, such as large-scale engineering manipulation or control of teleoperated robots. Docking haptics might also enable new haptics applications by enabling novel configurations of devices (c.f. Figure \ref{fig:cartoons}(g)-(i)).

The design space of docking haptics systems is very large and requires considerable work in various areas, but by designing with docking haptics in mind, alternative trade offs could be made leading to devices that better inter-operate for more flexible and general purpose haptic feedback in VR.

\section*{Acknowledgements}
This work was part-funded by UK EPSRC project Context Aware network architectures for Sending Multiple Senses (EP/P004016/1).

\clearpage

\bibliographystyle{abbrv-doi}
\bibliography{DockingHaptics}

\begin{thebibliography}{10}

\bibitem{Agah2004}
F.~Agah, M.~Mehrandezh, R.~G. Fenton, and B.~Benhabib.
\newblock {On-line robotic interception planning using a rendezvous-guidance
  technique}.
\newblock {\em Journal of Intelligent and Robotic Systems: Theory and
  Applications}, 40(1):23--44, 2004. doi: {{%
10\hspace{.1pt}\discretionary{.}{%
}{.}\hspace{.4pt}1023\discretionary{/}{%
}{/}B\discretionary{:}{%
}{:}JINT\hspace{.1pt}\discretionary{.}{%
}{.}\hspace{.4pt}0000034337\hspace{.1pt}\discretionary{.}{%
}{.}\hspace{.4pt}95125\hspace{.1pt}\discretionary{.}{%
}{.}\hspace{.4pt}bf}}


\bibitem{araujo_snake_2016}
B.~Araujo, R.~Jota, V.~Perumal, J.~X. Yao, K.~Singh, and D.~Wigdor.
\newblock Snake {Charmer}: {Physically} {Enabling} {Virtual} {Objects}.
\newblock In {\em Proceedings of the {TEI} '16: {Tenth} {International}
  {Conference} on {Tangible}, {Embedded}, and {Embodied} {Interaction} - {TEI}
  '16}, pp. 218--226. ACM Press, New York, New York, USA, 2016. doi: {{%
10\hspace{.1pt}\discretionary{.}{%
}{.}\hspace{.4pt}1145\discretionary{/}{%
}{/}2839462\hspace{.1pt}\discretionary{.}{%
}{.}\hspace{.4pt}2839484}}


\bibitem{azmandian_haptic_2016}
M.~Azmandian, M.~Hancock, H.~Benko, E.~Ofek, and A.~D. Wilson.
\newblock Haptic {Retargeting}: {Dynamic} {Repurposing} of {Passive} {Haptics}
  for {Enhanced} {Virtual} {Reality} {Experiences}.
\newblock In {\em Proceedings of the 2016 {CHI} {Conference} on {Human}
  {Factors} in {Computing} {Systems}}, {CHI} '16, pp. 1968--1979. ACM, New
  York, NY, USA, 2016. doi: {{%
10\hspace{.1pt}\discretionary{.}{%
}{.}\hspace{.4pt}1145\discretionary{/}{%
}{/}2858036\hspace{.1pt}\discretionary{.}{%
}{.}\hspace{.4pt}2858226}}


\bibitem{Bender2012}
J.~Bender, M.~M{\"{u}}ller, M.~A. Otaduy, and M.~Teschner.
\newblock {Position-based Methods for the Simulation of Solid Objects in
  Computer Graphics}.
\newblock In {\em STAR Proceedings of Eurographics 2013}, 2013. doi: {{%
10\hspace{.1pt}\discretionary{.}{%
}{.}\hspace{.4pt}2312\discretionary{/}{%
}{/}conf\discretionary{/}{%
}{/}EG2013\discretionary{/}{%
}{/}stars\discretionary{/}{%
}{/}001\discretionary{%
}{-}{-}022}}


\bibitem{benko_normaltouch_2016}
H.~Benko, C.~Holz, M.~Sinclair, and E.~Ofek.
\newblock {NormalTouch} and {TextureTouch}: {High}-fidelity 3d {Haptic} {Shape}
  {Rendering} on {Handheld} {Virtual} {Reality} {Controllers}.
\newblock In {\em Proceedings of the 29th {Annual} {Symposium} on {User}
  {Interface} {Software} and {Technology} - {UIST} '16}, pp. 717--728. ACM
  Press, New York, New York, USA, 2016. doi: {{%
10\hspace{.1pt}\discretionary{.}{%
}{.}\hspace{.4pt}1145\discretionary{/}{%
}{/}2984511\hspace{.1pt}\discretionary{.}{%
}{.}\hspace{.4pt}2984526}}


\bibitem{bergamasco_arm_1994}
M.~Bergamasco, B.~Allotta, L.~Bosio, L.~Ferretti, G.~Parrini, G.~M. Prisco,
  F.~Salsedo, and G.~Sartini.
\newblock An arm exoskeleton system for teleoperation and virtual environments
  applications.
\newblock In {\em Proceedings of the 1994 {IEEE} {International} {Conference}
  on {Robotics} and {Automation}}, vol.~2, pp. 1449--1454, May 1994. doi: {{%
10\hspace{.1pt}\discretionary{.}{%
}{.}\hspace{.4pt}1109\discretionary{/}{%
}{/}ROBOT\hspace{.1pt}\discretionary{.}{%
}{.}\hspace{.4pt}1994\hspace{.1pt}\discretionary{.}{%
}{.}\hspace{.4pt}351286}}


\bibitem{Borg2002}
J.~M. Borg, M.~Mehrandezh, R.~G. Fenton, and B.~Benhabib.
\newblock {Navigation-guidance-based robotic interception of moving objects in
  industrial settings}.
\newblock {\em Journal of Intelligent and Robotic Systems: Theory and
  Applications}, 33(1):1--23, 2002. doi: {{%
10\hspace{.1pt}\discretionary{.}{%
}{.}\hspace{.4pt}1023\discretionary{/}{%
}{/}A\discretionary{:}{%
}{:}1014490704273}}


\bibitem{bos_structured_2016}
R.~A. Bos, C.~J. Haarman, T.~Stortelder, K.~Nizamis, J.~L. Herder, A.~H.
  Stienen, and D.~H. Plettenburg.
\newblock A structured overview of trends and technologies used in dynamic hand
  orthoses.
\newblock {\em Journal of NeuroEngineering and Rehabilitation}, 13(62), Dec.
  2016. doi: {{%
10\hspace{.1pt}\discretionary{.}{%
}{.}\hspace{.4pt}1186\discretionary{/}{%
}{/}s12984\discretionary{%
}{-}{-}016\discretionary{%
}{-}{-}0168\discretionary{%
}{-}{-}z}}


\bibitem{Buchholz1992}
B.~Buchholz and T.~J. Armstrong.
\newblock {A kinematic model of the human hand to evaluate its prehensile
  capabilities}.
\newblock {\em Journal of Biomechanics}, 25(2):149--162, 1992.

\bibitem{burdea_force_1996}
G.~C. Burdea.
\newblock {\em Force and {Touch} {Feedback} for {Virtual} {Reality}}.
\newblock Wiley-Blackwell, New York, Sept. 1996.

\bibitem{choi_grabity:_2017}
I.~Choi, H.~Culbertson, M.~R. Miller, A.~Olwal, and S.~Follmer.
\newblock Grabity: {A} {Wearable} {Haptic} {Interface} for {Simulating}
  {Weight} and {Grasping} in {Virtual} {Reality}.
\newblock In {\em Proceedings of the 30th {Annual} {ACM} {Symposium} on {User}
  {Interface} {Software} and {Technology} - {UIST} '17}, pp. 119--130. ACM
  Press, New York, New York, USA, 2017. doi: {{%
10\hspace{.1pt}\discretionary{.}{%
}{.}\hspace{.4pt}1145\discretionary{/}{%
}{/}3126594\hspace{.1pt}\discretionary{.}{%
}{.}\hspace{.4pt}3126599}}


\bibitem{choi_wolverine:_2016}
I.~Choi and S.~Follmer.
\newblock Wolverine: {A} {Wearable} {Haptic} {Interface} for {Grasping} in
  {VR}.
\newblock In {\em Proceedings of the 29th {Annual} {Symposium} on {User}
  {Interface} {Software} and {Technology} - {UIST} '16 {Adjunct}}, pp.
  117--119. ACM Press, New York, New York, USA, 2016. doi: {{%
10\hspace{.1pt}\discretionary{.}{%
}{.}\hspace{.4pt}1145\discretionary{/}{%
}{/}2984751\hspace{.1pt}\discretionary{.}{%
}{.}\hspace{.4pt}2985725}}


\bibitem{Chwa2005}
D.~Chwa, J.~Kang, and J.~Y. Choi.
\newblock {Online trajectory planning of robot arms for interception of fast
  maneuvering object under torque and velocity constraints}.
\newblock {\em IEEE Transactions on Systems, Man, and Cybernetics Part
  A:Systems and Humans}, 35(6):831--843, 2005. doi: {{%
10\hspace{.1pt}\discretionary{.}{%
}{.}\hspace{.4pt}1109\discretionary{/}{%
}{/}TSMCA\hspace{.1pt}\discretionary{.}{%
}{.}\hspace{.4pt}2005\hspace{.1pt}\discretionary{.}{%
}{.}\hspace{.4pt}851340}}


\bibitem{Cobos2010}
S.~Cobos, M.~Ferre, and R.~Aracil.
\newblock {Simplified human hand models based on grasping analysis}.
\newblock {\em IEEE/RSJ 2010 International Conference on Intelligent Robots and
  Systems, IROS 2010 - Conference Proceedings}, pp. 610--615, 2010. doi: {{%
10\hspace{.1pt}\discretionary{.}{%
}{.}\hspace{.4pt}1109\discretionary{/}{%
}{/}IROS\hspace{.1pt}\discretionary{.}{%
}{.}\hspace{.4pt}2010\hspace{.1pt}\discretionary{.}{%
}{.}\hspace{.4pt}5651479}}


\bibitem{Cobos2008}
S.~Cobos, M.~Ferre, M.~Sanchez~Uran, J.~Ortego, and C.~Pena.
\newblock {Efficient human hand kinematics for manipulation tasks}.
\newblock In {\em 2008 IEEE/RSJ International Conference on Intelligent Robots
  and Systems}, pp. 2246--2251. IEEE, 9 2008. doi: {{%
10\hspace{.1pt}\discretionary{.}{%
}{.}\hspace{.4pt}1109\discretionary{/}{%
}{/}IROS\hspace{.1pt}\discretionary{.}{%
}{.}\hspace{.4pt}2008\hspace{.1pt}\discretionary{.}{%
}{.}\hspace{.4pt}4651053}}


\bibitem{coelho_shape-changing_2011}
M.~Coelho and J.~Zigelbaum.
\newblock Shape-changing interfaces.
\newblock {\em Personal and Ubiquitous Computing}, 15(2):161--173, 2011. doi:
  {{%
10\hspace{.1pt}\discretionary{.}{%
}{.}\hspace{.4pt}1007\discretionary{/}{%
}{/}s00779\discretionary{%
}{-}{-}010\discretionary{%
}{-}{-}0311\discretionary{%
}{-}{-}y}}


\bibitem{culbertson_haptics:_2018}
H.~Culbertson, S.~B. Schorr, and A.~M. Okamura.
\newblock Haptics: {The} {Present} and {Future} of {Artificial} {Touch}
  {Sensation}.
\newblock {\em Annual Review of Control, Robotics, and Autonomous Systems},
  1:385--409, 2018. doi: {{%
10\hspace{.1pt}\discretionary{.}{%
}{.}\hspace{.4pt}1146\discretionary{/}{%
}{/}annurev\discretionary{%
}{-}{-}control\discretionary{%
}{-}{-}060117\discretionary{%
}{-}{-}105043}}


\bibitem{cyberglove_systems_llc_cyberforce_nodate}
{CyberGlove Systems LLC}.
\newblock {CyberForce}, 2019.
\newblock Available at: http://www.cyberglovesystems.com/cyberforce [Accessed
  November 11, 2019].

\bibitem{cyberglove_systems_llc_cybergrasp_nodate}
{CyberGlove Systems LLC}.
\newblock {CyberGrasp}, 2019.
\newblock Available at: http://www.cyberglovesystems.com/cybergrasp [Accessed
  November 11, 2019].

\bibitem{endo_five-fingered_2009}
T.~Endo, H.~Kawasaki, T.~Mouri, Y.~Doi, T.~Yoshida, Y.~Ishigure, H.~Shimomura,
  M.~Matsumura, and K.~Koketsu.
\newblock Five-fingered haptic interface robot: {HIRO} {III}.
\newblock In {\em World {Haptics} 2009 - {Third} {Joint} {EuroHaptics}
  conference and {Symposium} on {Haptic} {Interfaces} for {Virtual}
  {Environment} and {Teleoperator} {Systems}}, pp. 458--463. IEEE, Mar. 2009.
  doi: {{%
10\hspace{.1pt}\discretionary{.}{%
}{.}\hspace{.4pt}1109\discretionary{/}{%
}{/}WHC\hspace{.1pt}\discretionary{.}{%
}{.}\hspace{.4pt}2009\hspace{.1pt}\discretionary{.}{%
}{.}\hspace{.4pt}4810812}}


\bibitem{fang_wireality_2020}
C.~Fang, Y.~Zhang, M.~Dworman, and C.~Harrison.
\newblock Wireality: {Enabling} {Complex} {Tangible} {Geometries} in {Virtual}
  {Reality} with {Worn} {Multi}-{String} {Haptics}.
\newblock In {\em Proceedings of the 2020 {CHI} {Conference} on {Human}
  {Factors} in {Computing} {Systems}}, {CHI} '20, pp. 1--10. Association for
  Computing Machinery, Honolulu, HI, USA, Apr. 2020. doi: {{%
10\hspace{.1pt}\discretionary{.}{%
}{.}\hspace{.4pt}1145\discretionary{/}{%
}{/}3313831\hspace{.1pt}\discretionary{.}{%
}{.}\hspace{.4pt}3376470}}


\bibitem{galiana_multi-finger_2015}
I.~Galiana and M.~Ferre, eds.
\newblock {\em Multi-finger {Haptic} {Interaction}}.
\newblock Springer, London, 2013 edition ed., June 2013.

\bibitem{langrana_integration_1995}
D.~Gomez, G.~Burdea, and N.~Langrana.
\newblock Integration of the {Rutgers} {Master} {II} in a virtual reality
  simulation.
\newblock In {\em Proceedings Virtual Reality Annual International Symposium
  '95}, pp. 198--202, 1995. doi: {{%
10\hspace{.1pt}\discretionary{.}{%
}{.}\hspace{.4pt}1109\discretionary{/}{%
}{/}VRAIS\hspace{.1pt}\discretionary{.}{%
}{.}\hspace{.4pt}1995\hspace{.1pt}\discretionary{.}{%
}{.}\hspace{.4pt}512496}}


\bibitem{gopura_developments_2016}
R.~A. R.~C. Gopura, D.~S.~V. Bandara, K.~Kiguchi, and G.~K.~I. Mann.
\newblock Developments in hardware systems of active upper-limb exoskeleton
  robots: {A} review.
\newblock {\em Robotics and Autonomous Systems}, 75:203--220, Jan. 2016. doi:
  {{%
10\hspace{.1pt}\discretionary{.}{%
}{.}\hspace{.4pt}1016\discretionary{/}{%
}{/}j\hspace{.1pt}\discretionary{.}{%
}{.}\hspace{.4pt}robot\hspace{.1pt}\discretionary{.}{%
}{.}\hspace{.4pt}2015\hspace{.1pt}\discretionary{.}{%
}{.}\hspace{.4pt}10\hspace{.1pt}\discretionary{.}{%
}{.}\hspace{.4pt}001}}


\bibitem{gosselin_widening_2007}
F.~Gosselin, C.~Andriot, F.~Bergez, and X.~Merlhiot.
\newblock Widening 6-{DOF} haptic devices workspace with an additional degree
  of freedom.
\newblock In {\em 2007 2nd {Joint} {EuroHaptics} {Conference} and {Symposium}
  on {Haptic} {Interfaces} for {Virtual} {Environments} and {Teleoperator}
  {Systems}({WHC})}, pp. 452--457, Mar. 2007. doi: {{%
10\hspace{.1pt}\discretionary{.}{%
}{.}\hspace{.4pt}1109\discretionary{/}{%
}{/}WHC\hspace{.1pt}\discretionary{.}{%
}{.}\hspace{.4pt}2007\hspace{.1pt}\discretionary{.}{%
}{.}\hspace{.4pt}127}}


\bibitem{gosselin_large_2008}
F.~Gosselin, C.~Andriot, J.~Savall, and J.~Martín.
\newblock Large {Workspace} {Haptic} {Devices} for {Human}-{Scale}
  {Interaction}: {A} {Survey}.
\newblock In M.~Ferre, ed., {\em Haptics: {Perception}, {Devices} and
  {Scenarios}}, Lecture {Notes} in {Computer} {Science}, pp. 523--528. Springer
  Berlin Heidelberg, 2008.

\bibitem{Gu2016}
X.~Gu, Y.~Zhang, W.~Sun, Y.~Bian, D.~Zhou, and P.~O. Kristensson.
\newblock {Dexmo: An Inexpensive and Lightweight Mechanical Exoskeleton for
  Motion Capture and Force Feedback in VR}.
\newblock {\em Proceedings of the 2016 CHI Conference on Human Factors in
  Computing Systems - CHI '16}, pp. 1991--1995, 2016. doi: {{%
10\hspace{.1pt}\discretionary{.}{%
}{.}\hspace{.4pt}1145\discretionary{/}{%
}{/}2858036\hspace{.1pt}\discretionary{.}{%
}{.}\hspace{.4pt}2858487}}


\bibitem{haption_inca_nodate}
Haption.
\newblock Inca\textsuperscript{TM}, 2019.
\newblock Available at: https://www.haption.com/en/products-en/inca-en.html
  [Accessed November 11, 2019].

\bibitem{haption_scale1_nodate}
Haption.
\newblock Scale1\textsuperscript{TM}, 2019.
\newblock Available at:
  https://www.haption.com/en/products-en/scale-one-en.html [Accessed November
  11, 2019].

\bibitem{haption_virtuose_nodate}
{Haption}.
\newblock Virtuose\textsuperscript{TM} 6d, 2019.
\newblock Available at:
  https://www.haption.com/en/products-en/virtuose-6d-en.html [Accessed November
  11, 2019].

\bibitem{hartenberg1964kinematic}
R.~S. Hartenberg and J.~Denavit.
\newblock {\em Kinematic synthesis of linkages}.
\newblock McGraw-Hill, New York, 1964.

\bibitem{heo_current_2012}
P.~Heo, G.~M. Gu, S.-j. Lee, K.~Rhee, and J.~Kim.
\newblock Current hand exoskeleton technologies for rehabilitation and
  assistive engineering.
\newblock {\em International Journal of Precision Engineering and
  Manufacturing}, 13(5):807--824, May 2012. doi: {{%
10\hspace{.1pt}\discretionary{.}{%
}{.}\hspace{.4pt}1007\discretionary{/}{%
}{/}s12541\discretionary{%
}{-}{-}012\discretionary{%
}{-}{-}0107\discretionary{%
}{-}{-}2}}


\bibitem{heo_thors_2018}
S.~Heo, C.~Chung, G.~Lee, and D.~Wigdor.
\newblock Thor's {Hammer}: {An} {Ungrounded} {Force} {Feedback} {Device}
  {Utilizing} {Propeller}-{Induced} {Propulsive} {Force}.
\newblock In {\em Proceedings of the 2018 {CHI} {Conference} on {Human}
  {Factors} in {Computing} {Systems}}, {CHI} '18, pp. 525:1--525:11. ACM, New
  York, NY, USA, 2018. doi: {{%
10\hspace{.1pt}\discretionary{.}{%
}{.}\hspace{.4pt}1145\discretionary{/}{%
}{/}3173574\hspace{.1pt}\discretionary{.}{%
}{.}\hspace{.4pt}3174099}}


\bibitem{hinckley_passive_1994}
K.~Hinckley, R.~Pausch, J.~C. Goble, and N.~F. Kassell.
\newblock Passive {Real}-world {Interface} {Props} for {Neurosurgical}
  {Visualization}.
\newblock In {\em Proceedings of the {SIGCHI} {Conference} on {Human} {Factors}
  in {Computing} {Systems}}, {CHI} '94, pp. 452--458. ACM, New York, NY, USA,
  1994. doi: {{%
10\hspace{.1pt}\discretionary{.}{%
}{.}\hspace{.4pt}1145\discretionary{/}{%
}{/}191666\hspace{.1pt}\discretionary{.}{%
}{.}\hspace{.4pt}191821}}


\bibitem{Hrabia2013}
C.-E. Hrabia, K.~Wolf, and M.~Wilhelm.
\newblock {Whole Hand Modeling using 8 Wearable Sensors: Biomechanics for Hand
  Pose Prediction}.
\newblock In {\em Proceedings of the 4th Augmented Human International
  Conference}, pp. 21--28, 2013. doi: {{%
10\hspace{.1pt}\discretionary{.}{%
}{.}\hspace{.4pt}1145\discretionary{/}{%
}{/}2459236\hspace{.1pt}\discretionary{.}{%
}{.}\hspace{.4pt}2459241}}


\bibitem{insko_passive_2001}
B.~E. Insko.
\newblock {\em Passive {Haptics} {Significantly} {Enhances} {Virtual}
  {Environments}}.
\newblock {PhD} {Thesis}, The University of North Carolina at Chapel Hill,
  2001.

\bibitem{Kunwar2006}
F.~Kunwar, F.~Wong, R.~B. Mrad, and B.~Benhabib.
\newblock {Guidance-based on-line robot motion planning for the interception of
  mobile targets in dynamic environments}.
\newblock {\em Journal of Intelligent and Robotic Systems: Theory and
  Applications}, 47(4):341--360, 2006. doi: {{%
10\hspace{.1pt}\discretionary{.}{%
}{.}\hspace{.4pt}1007\discretionary{/}{%
}{/}s10846\discretionary{%
}{-}{-}006\discretionary{%
}{-}{-}9080\discretionary{%
}{-}{-}2}}


\bibitem{Li2007a}
Z.~M. Li and J.~Tang.
\newblock {Coordination of thumb joints during opposition}.
\newblock {\em Journal of Biomechanics}, 40(3):502--510, 2007.

\bibitem{Lin1989}
Z.~Lin, V.~Zeman, and R.~Patel.
\newblock {On-line robot trajectory planning for catching a moving object}.
\newblock In {\em Proceedings, 1989 International Conference on Robotics and
  Automation}, pp. 1726--1731, 1989.

\bibitem{ma_rml_2015}
Z.~MA and P.~Ben-Tzvi.
\newblock Rml glove - an exoskeleton glove mechanism with haptics feedback.
\newblock {\em IEEE/ASME Transactions on Mechatronics}, 20(2):641--652, Apr.
  2015. doi: {{%
10\hspace{.1pt}\discretionary{.}{%
}{.}\hspace{.4pt}1109\discretionary{/}{%
}{/}TMECH\hspace{.1pt}\discretionary{.}{%
}{.}\hspace{.4pt}2014\hspace{.1pt}\discretionary{.}{%
}{.}\hspace{.4pt}2305842}}


\bibitem{massie_phantom_1994}
T.~H. Massie and J.~K. Salisbury.
\newblock The phantom haptic interface: {A} device for probing virtual objects.
\newblock In {\em Proceedings of the {ASME} winter annual meeting, symposium on
  haptic interfaces for virtual environment and teleoperator systems}, vol.~55,
  pp. 295--300. Citeseer, 1994.

\bibitem{mcneely_robotic_1993}
W.~A. McNeely.
\newblock Robotic graphics: a new approach to force feedback for virtual
  reality.
\newblock In {\em Proceedings of {IEEE} {Virtual} {Reality} {Annual}
  {International} {Symposium}}, pp. 336--341, Sept. 1993.

\bibitem{Mehrandezh2000}
M.~Mehrandezh, N.~M. Sela, R.~G. Fenton, and B.~Benhabib.
\newblock {Robotic Interception of Moving Objects Using an Augmented Ideal
  Proportional Navigation Guidance Technique}.
\newblock {\em IEEE Transactions on Systems, Man, and Cybernetics Part
  A:Systems and Humans}, 30(3):238--250, 2000.

\bibitem{ott_two-handed_2010}
R.~Ott, F.~Vexo, and D.~Thalmann.
\newblock Two-handed {Haptic} {Manipulation} for {CAD} and {VR} {Applications}.
\newblock {\em Computer-Aided Design and Applications}, 7(1):125--138, Jan.
  2010. doi: {{%
10\hspace{.1pt}\discretionary{.}{%
}{.}\hspace{.4pt}3722\discretionary{/}{%
}{/}cadaps\hspace{.1pt}\discretionary{.}{%
}{.}\hspace{.4pt}2010\hspace{.1pt}\discretionary{.}{%
}{.}\hspace{.4pt}125\discretionary{%
}{-}{-}138}}


\bibitem{pavlik_interacting_2013}
R.~A. Pavlik, J.~M. Vance, and G.~R. Luecke.
\newblock Interacting {With} a {Large} {Virtual} {Environment} by {Combining} a
  {Ground}-{Based} {Haptic} {Device} and a {Mobile} {Robot} {Base}.
\newblock In {\em Proceedings of the 33rd Computers and Information in
  Engineering Conference}, vol.~2B, Aug. 2013. doi: {{%
10\hspace{.1pt}\discretionary{.}{%
}{.}\hspace{.4pt}1115\discretionary{/}{%
}{/}DETC2013\discretionary{%
}{-}{-}13441}}


\bibitem{peer_new_2008}
A.~Peer and M.~Buss.
\newblock A {New} {Admittance}-{Type} {Haptic} {Interface} for {Bimanual}
  {Manipulations}.
\newblock {\em IEEE/ASME Transactions on Mechatronics}, 13(4):416--428, Aug.
  2008.

\bibitem{sato_spidar_2002}
M.~Sato.
\newblock {SPIDAR} and virtual reality.
\newblock In {\em Proceedings of the 5th {Biannual} {World} {Automation}
  {Congress}}, vol.~13, pp. 17--23, June 2002.

\bibitem{Springer2002}
S.~L. Springer and N.~J. Ferrier.
\newblock {Design and Control of a Force-Reflecting Haptic Interface for
  Teleoperational Grasping}.
\newblock {\em Journal of Mechanical Design}, 124(2):277--283, 2002.

\bibitem{sturdee_analysis_2018}
M.~Sturdee and J.~Alexander.
\newblock Analysis and {Classification} of {Shape}-{Changing} {Interfaces} for
  {Design} and {Application}-based {Research}.
\newblock {\em ACM Computing Surveys}, 51(1):1--32, 2018. doi: {{%
10\hspace{.1pt}\discretionary{.}{%
}{.}\hspace{.4pt}1145\discretionary{/}{%
}{/}3143559}}


\bibitem{talvas_survey_2014}
A.~Talvas, M.~Marchal, and A.~L\'ecuyer.
\newblock A {Survey} on {Bimanual} {Haptic} {Interaction}.
\newblock {\em IEEE Transactions on Haptics}, 7(3):285--300, July 2014. doi:
  {{%
10\hspace{.1pt}\discretionary{.}{%
}{.}\hspace{.4pt}1109\discretionary{/}{%
}{/}TOH\hspace{.1pt}\discretionary{.}{%
}{.}\hspace{.4pt}2014\hspace{.1pt}\discretionary{.}{%
}{.}\hspace{.4pt}2314456}}


\bibitem{todorov_mujoco:_2012}
E.~Todorov, T.~Erez, and Y.~Tassa.
\newblock {MuJoCo}: {A} physics engine for model-based control.
\newblock In {\em 2012 {IEEE}/{RSJ} {International} {Conference} on
  {Intelligent} {Robots} and {Systems}}, pp. 5026--5033, Oct. 2012. doi: {{%
10\hspace{.1pt}\discretionary{.}{%
}{.}\hspace{.4pt}1109\discretionary{/}{%
}{/}IROS\hspace{.1pt}\discretionary{.}{%
}{.}\hspace{.4pt}2012\hspace{.1pt}\discretionary{.}{%
}{.}\hspace{.4pt}6386109}}


\bibitem{ueberle_vishard10_2004}
M.~Ueberle, N.~Mock, and M.~Buss.
\newblock {VISHARD}10, a novel hyper-redundant haptic interface.
\newblock In {\em 12th {International} {Symposium} on {Haptic} {Interfaces} for
  {Virtual} {Environment} and {Teleoperator} {Systems}, 2004. {HAPTICS} '04.
  {Proceedings}.}, pp. 58--65, Mar. 2004. doi: {{%
10\hspace{.1pt}\discretionary{.}{%
}{.}\hspace{.4pt}1109\discretionary{/}{%
}{/}HAPTIC\hspace{.1pt}\discretionary{.}{%
}{.}\hspace{.4pt}2004\hspace{.1pt}\discretionary{.}{%
}{.}\hspace{.4pt}1287178}}


\bibitem{VanDerHulst2012}
F.~P. Van Der~Hulst, S.~Sch{\"{a}}tzle, C.~Preusche, and A.~Schiele.
\newblock {A functional anatomy based kinematic human hand model with simple
  size adaptation}.
\newblock {\em Proceedings - IEEE International Conference on Robotics and
  Automation}, pp. 5123--5129, 2012. doi: {{%
10\hspace{.1pt}\discretionary{.}{%
}{.}\hspace{.4pt}1109\discretionary{/}{%
}{/}ICRA\hspace{.1pt}\discretionary{.}{%
}{.}\hspace{.4pt}2012\hspace{.1pt}\discretionary{.}{%
}{.}\hspace{.4pt}6225350}}


\bibitem{lammertse_hapticmaster_2002}
R.~Van~der Linde, P.~Lammertse, E.~Frederiksen, and B.~Ruiter.
\newblock The {HapticMaster}, a new high-performance haptic interface.
\newblock In {\em Proceedings of Eurohaptics}, 2002.

\bibitem{whitmire_haptic_2018}
E.~Whitmire, H.~Benko, C.~Holz, E.~Ofek, and M.~Sinclair.
\newblock Haptic {Revolver}: {Touch}, {Shear}, {Texture}, and {Shape}
  {Rendering} on a {Reconfigurable} {Virtual} {Reality} {Controller}.
\newblock In {\em Proceedings of the 2018 {CHI} {Conference} on {Human}
  {Factors} in {Computing} {Systems}}, {CHI} '18, pp. 86:1--86:12. ACM, New
  York, NY, USA, 2018. doi: {{%
10\hspace{.1pt}\discretionary{.}{%
}{.}\hspace{.4pt}1145\discretionary{/}{%
}{/}3173574\hspace{.1pt}\discretionary{.}{%
}{.}\hspace{.4pt}3173660}}


\bibitem{zenner_shifty:_2017}
A.~Zenner and A.~Kr{\"u}ger.
\newblock Shifty: {A} {Weight}-{Shifting} {Dynamic} {Passive} {Haptic} {Proxy}
  to {Enhance} {Object} {Perception} in {Virtual} {Reality}.
\newblock {\em IEEE Transactions on Visualization and Computer Graphics},
  23(4):1285--1294, Apr. 2017. doi: {{%
10\hspace{.1pt}\discretionary{.}{%
}{.}\hspace{.4pt}1109\discretionary{/}{%
}{/}TVCG\hspace{.1pt}\discretionary{.}{%
}{.}\hspace{.4pt}2017\hspace{.1pt}\discretionary{.}{%
}{.}\hspace{.4pt}2656978}}


\bibitem{zhao_robotic_2017}
Y.~Zhao, L.~H. Kim, Y.~Wang, M.~Le~Goc, and S.~Follmer.
\newblock Robotic {Assembly} of {Haptic} {Proxy} {Objects} for {Tangible}
  {Interaction} and {Virtual} {Reality}.
\newblock In {\em Proceedings of the 2017 {ACM} {International} {Conference} on
  {Interactive} {Surfaces} and {Spaces}}, {ISS} '17, pp. 82--91. ACM, New York,
  NY, USA, 2017. doi: {{%
10\hspace{.1pt}\discretionary{.}{%
}{.}\hspace{.4pt}1145\discretionary{/}{%
}{/}3132272\hspace{.1pt}\discretionary{.}{%
}{.}\hspace{.4pt}3134143}}


\end{thebibliography}

\end{document}